\documentclass[12pt]{article}
\usepackage[english,activeacute]{babel}
\usepackage{natbib}
\usepackage{comment}
\usepackage{float}
\usepackage[hidelinks]{hyperref}
\usepackage{mathrsfs}
\usepackage{enumitem}
\usepackage[font={small,it}]{caption}
\usepackage{amsmath,amsfonts,amsthm,amssymb}
\usepackage{bm,rotating,multirow,dsfont,graphicx}
\usepackage[usenames, dvipsnames]{color}
\usepackage{url}
\usepackage{multicol}
\usepackage{multirow}
\usepackage[T1]{fontenc}
\usepackage{flafter}
\usepackage{appendix}
\usepackage{subfigure}
\usepackage{xcolor}
\usepackage{soul}
\usepackage{setspace}
\usepackage{booktabs}
\makeatletter
\def\hlinewd#1{%
	\noalign{\ifnum0=`}\fi\hrule \@height #1 %
	\futurelet\reserved@a\@xhline}
\makeatother
\def\spacingset#1{\renewcommand{\baselinestretch}{#1}\small\normalsize}\spacingset{1}
\def\@roman#1{\romannumeral #1}

\begin{document}

\title{Hybrid Bayesian Models for Community Detection with Application to a Colombian Conflict Network}

\date{}

\author{
    Juan Sosa, Universidad Nacional de Colombia, Colombia\footnote{Corresponding author: jcsosam@unal.edu.co.} \\
    Eleni Dilma, University of Florida, US \\
    Brenda Betancourt, NORC at the University of Chicago, US \\
}

\maketitle


\begin{abstract}
We introduce a flexible Bayesian framework for clustering nodes in undirected binary networks, motivated by the need to uncover structural patterns in complex environments. Building on the stochastic block model, we develop two hybrid extensions: the Class-Distance Model, which governs interaction probabilities through Euclidean distances between cluster-level latent positions, and the Class-Bilinear Model, which captures more complex relational patterns via bilinear interactions. We apply this framework to a novel network derived from the Colombian armed conflict, where municipalities are connected through the co-presence of armed actors, violence, and illicit economies. The resulting clusters align with empirical patterns of territorial control and trafficking corridors, highlighting the models' capacity to recover and explain complex  dynamics. Full Bayesian inference is carried out via MCMC under both finite and nonparametric clustering priors. While the main application centers on the Colombian conflict, we also assess model performance using synthetic data as well as other two benchmark datasets.
\end{abstract}


\spacingset{1.1} 

\newpage

\section{Introduction}

The study of information emerging from the interconnectedness of autonomous elements in a system is crucial for understanding diverse phenomena. The structure formed by these elements (individuals or actors) and their interactions (ties or connections) is commonly referred to as a \textit{network}. Networks are prevalent across various disciplines, including finance, social sciences, biology, epidemiology, and computer science, among others. The diversity of entities and connections within networks highlights their broad applicability and importance in research.

Statistical methods have substantially advanced to enhance our understanding of how the attributes and relationships of actors shape network behavior. These methods serve three primary purposes: describing individual entity characteristics and relational patterns, constructing stochastic models to explain network formation processes, and predicting missing or future connections based on local interaction rules and structural properties. Unlike deterministic approaches, statistical models incorporate measures of uncertainty, providing robust estimates and predictions.

A central framework in probabilistic network modeling is the stochastic block model (SBM) and its extensions. Recent developments have focused on relaxing the rigid assumptions of traditional SBMs to capture more complex interaction structures and latent heterogeneity. For instance, \citet{legramanti2022extended} propose an extended SBM where connection probabilities are modulated by latent distances between cluster-specific latent vectors, unifying ideas from distance-based and block-based models to accommodate overlapping community structure and hierarchical patterns. Similarly, the Covariate-Assisted Latent Factor SBM (CALF-SBM) of \citet{louit2025calf} introduces a bilinear structure where edge probabilities are influenced by both latent node positions and observed covariates, enabling more flexible modeling of dependence structures while maintaining interpretability and scalability.

Building upon this recent literature, our contribution is the introduction of a hybrid class of models that combines block interaction with latent structure at the group level, allowing interactions to depend not only on cluster membership, but also on low-dimensional latent features associated with each group. This reparameterization improves model fitting and predictive capabilities by incorporating geometric constraints (e.g., through latent distances or bilinear inner products), while reducing the complexity of modeling all pairwise interactions. The proposed framework offers greater flexibility than traditional SBMs while avoiding the overparameterization commonly found in full latent space models.  For a detailed discussion of node characterization, network structure, and community detection, see \citet{kolaczyk-2020}, \citet{newman-2018}, and \citet{barabasi-2016}. Furthermore, for a comprehensive treatment of cross-sectional network modeling, refer to \citet{goldenberg-2010}, \citet{snijders-2011}, and \citet{crane-2018-probabilistic}.

The structure of the paper is as follows. Section~\ref{sec:related_work} provides a formal background on stochastic block models and their Bayesian variants, including the extended class of latent structure SBM models. Section~\ref{sec:hybrid} introduces our proposed class-distance and class-bilinear formulations, accounting for modeling fitting, hyperparameter elicitation and model evaluation. Section~\ref{sec:simulation} contains simulation studies illustrating the behavior of the proposed models under different settings. Section~\ref{sec:illustration} applies our models to real-world networks and compares performance with existing methods. Finally, Section~\ref{sec:discussion} concludes with a discussion and future research directions.

\section{Related work}\label{sec:related_work}

This section discusses the core concepts and existing models that serve as the basis for our proposed modeling approach. The focus of our work is on models for undirected binary networks, where connections are symmetric and represented as a dichotomous variable indicating presence or absence.

\subsection{Community detection}

One of the most well-studied problems in network analysis is community detection, which involves partitioning a graph into communities of nodes, also known as clusters or segments. Formally, a partition \( \mathcal{C} = \{C_1, \dots, C_K\} \) of the finite set of vertices \( V \) decomposes \( V \) into \( K \) disjoint, nonempty subsets \( C_k \), such that \( V = \bigcup_{k=1}^{K} C_k \). The goal of community detection in an assortative manner is to identify cohesive subgroups within a network, where nodes within the same group exhibit strong connections, while interactions between groups are relatively sparse. This has led to the development of a broad range of algorithmic and probabilistic methods. Understanding these structures is essential for various applications, including financial markets, social networks, information networks, cybersecurity, epidemiology, neuroscience, biological systems, transportation systems, and recommendation systems. For detailed reviews on community detection approaches, see \citet{newman-2004}, \citet{schaeffer-2007}, \citet{porter-2009}, \citet{lancichinetti2009community}, and \citet{fortunato-2010}.

Many of the most widely used approaches for community detection rely on deterministic algorithms, which partition a network into groups based on predefined heuristics. The most successful methods generally fall into two main categories: hierarchical clustering and spectral partitioning. Despite their widespread use, these methods have several limitations. A key drawback is their inability to quantify the uncertainty associated with the detected communities, as they do not account for variations that may arise due to noise or alternative interpretations of the network structure. In addition, most of these approaches are inherently designed to detect assortative structures, where nodes within the same community are more densely connected than those in different communities. However, this assumption does not always hold, as some networks exhibit disassortative mixing patterns, where nodes in different communities interact more frequently than those within the same group.

To address these challenges, probabilistic models provide a more flexible framework for community detection by incorporating stochastic processes to infer group memberships. A seminal family of models, introduced by \citet{holland1983stochastic}, is based on stochastic blockmodels (SBMs). SBMs explicitly define interaction probabilities between groups of nodes by assigning each node to a latent community and determining the probability of an edge between two nodes based on their respective group memberships. This approach is particularly appealing as it naturally accommodates both assortative and disassortative mixing patterns, making it a flexible tool for modeling diverse network structures. 

SBMs have been widely studied and extended in several directions. \citet{nowicki2001estimation} introduced a Bayesian framework for stochastic blockmodels, incorporating prior distributions to enhance inference. \citet{kemp2006learning} extended SBMs by allowing an infinite number of latent communities through a nonparametric Bayesian approach based on the Chinese restaurant process (e.g., \citealt{gershman2012tutorial}). \citet{airoldi2009mixed} proposed the mixed membership stochastic blockmodel, which allows nodes to belong to multiple communities with varying probabilities, capturing more complex relational structures. \citet{peixoto2014hierarchical} developed a hierarchical Bayesian formulation that improves model selection and parameter estimation for large-scale networks. More recently, advancements in variational inference techniques \citep{gopalan2013efficient} and spectral clustering methods \citep{rohe2011spectral} have enhanced the scalability and efficiency of SBM inference. Additionally, SBMs have been applied in various domains, including social networks, biological systems, and political science, as discussed in \citet{goldenberg-2010} and \citet{karrer2011stochastic}. See \citet{abbe2018community} for recent advancements in community detection and SBMs. Additionally, \citet{lee2019review} provides a comprehensive review of SBMs and their extensions for graph clustering.

\subsubsection{Stochastic block models}\label{sec_stochastic_blockmodels}

We focus on the analysis of undirected binary networks, which can be represented by the adjacency matrix \(\mathbf{Y} = [y_{i,j}]\), where \(1 \leq i < j \leq n\), and \(n\) denotes the total number of vertices in the network. The binary nature of interactions naturally suggests modeling them using a Bernoulli distribution: \( y_{i,j} \mid \theta_{i,j} \sim \textsf{Ber}(\theta_{i,j}) \). The core idea behind SBMs is that the network can be partitioned into \( K \leq n \) communities, where two vertices belong to the same community if they share the same interaction probabilities across the network. Formally, \( \theta_{i,j} = \Phi(\eta_{\phi(\xi_i, \xi_j)}) \), where the cluster assignments \( \xi_1, \dots, \xi_n \) take values in the set \(\{1, \dots, K\}\), the function \( \phi(x,y) = (\min(x,y),\max(x,y)) \) ensures a proper mapping over the set of parameters \(\eta_{k,\ell}\) for \( 1 \leq k \leq \ell \leq K \), which define the interaction structure between communities, and \( \Phi(\cdot) \) is the cumulative distribution function of a standard Normal random variable, which serves as the link function (other link functions can be considered).

Typically, the community parameters \( \eta_{k,\ell} \) are assumed to follow the prior distribution \( \eta_{k,\ell} \mid \zeta, \tau^2 \overset{\text{iid}}{\sim} \textsf{N}(\zeta, \tau^2) \), where \(\zeta\) represents the global mean interaction strength, and \(\tau^2\) governs the variability across communities. To complete the Bayesian specification, the hyperpriors are assigned as \( \zeta \sim \textsf{N}(\mu_\zeta, \sigma^2_\zeta) \) and \( \tau^2 \sim \textsf{IG}(a_\tau, b_\tau) \). This hierarchical structure allows the model to adaptively infer the community interaction effects while incorporating uncertainty in both the global mean and the variability of \( \eta_{k,\ell} \).

Assuming conditional independence of interactions both within and across actors, given \( \boldsymbol\eta = (\eta_{1,1},\ldots,\eta_{K,K}) \) and \( \boldsymbol{\xi} = (\xi_1,\ldots,\xi_n) \), the joint sampling distribution can be expressed as  
\begin{equation}\label{eq_class_likelihood}
p(\mathbf{Y} \mid \boldsymbol\eta, \boldsymbol\xi) 
= \prod_{k=1}^{K} \prod_{\ell=k}^{K} \left[\Phi(\eta_{k,\ell})\right]^{s_{k,\ell}} \left[1 - \Phi(\eta_{k,\ell})\right]^{n_{k,\ell} - s_{k,\ell}},
\end{equation} 
where \( K \) is the number of communities, \( s_{k,\ell} = \sum_{(i,j) \in S_{k,\ell}} y_{i,j} \), and \( n_{k,\ell} = \sum_{(i,j) \in S_{k,\ell}} 1 \), with summations taken over \( S_{k,\ell} = \{(i,j) : i < j, (k, \ell) = \phi(\xi_i, \xi_j) \} \). Additionally, note that actors \( i \) and \( j \) belong to the same class if and only if \( \xi_i = \xi_j \). A vector \( \boldsymbol{\xi} \) and its relabelings are treated as distinct objects, even though they correspond to the same partition. We assume that \( \boldsymbol{\xi} \) represents all possible relabelings that yield the same partition (swapping labels does not change the partition structure). At this point, it is important to distinguish between the total number of possible clusters $K$ and the number of occupied clusters $K^*$. The latter is defined as the number of distinct labels in \( \boldsymbol{\xi} \) and is upper bounded by \( \min(n, K) \).

One alternative to handling \( K \) is to fix it while intentionally overestimating the number of nonempty communities in the network, leading to a finite mixture model similar to that of \citet{nowicki2001estimation}. Under this formulation, known as the Dirichlet-Multinomial (DM) prior, \( \boldsymbol{\xi} \) is assumed to follow a Categorical distribution over \( \{1, \dots, K\} \), such that $\Pr(\xi_i = k \mid \omega_k) = \omega_k$, for $i=1,\ldots,n$, where \( \boldsymbol{\omega} = (\omega_1,\ldots,\omega_K) \) is a probability vector satisfying \( \sum_{k=1}^{K} \omega_k = 1 \), with $\boldsymbol{\omega} \mid \alpha \sim \textsf{Dir} \left(\alpha/K, \dots, \alpha/K \right)$. This formulation has a direct connection with the Dirichlet process (DP) prior as \( K \to \infty \), since after integrating out \( \boldsymbol{\omega} \), it follows that \( \boldsymbol{\xi} \) follows a Dirichlet-Multinomial distribution given by  
\[
p(\boldsymbol{\xi}) = \frac{\Gamma(\alpha)}{\Gamma(n + \alpha)} \prod_{k=1}^{K} \frac{\Gamma(n_k + \alpha/K)}{\Gamma(\alpha/K)},
\]
where \( n_k = \sum_{i=1}^n I(\xi_i = k) \) is the number of actors assigned to community \( k \) (see \citealt{neal2000markov} and \citealt{ishwaran2000markov}).

Alternatively, the infinite relational model of \citet{kemp2006learning} allows \( K^* \) to be inferred from the data using a nonparametric approach. Specifically, \( \boldsymbol{\xi} \) is assumed to follow a DP prior, which implies that its distribution follows the Ewens sampling formula:  
\[
p(\boldsymbol{\xi}) = \frac{\Gamma(\alpha)}{\Gamma(\alpha + n)}\, \alpha^{K^*} \prod_{k=1}^{K^*} \Gamma(n_k),
\]  
where $K^*$ is the number of nonempty communities in the network \citep{crane2016ubiquitous}. Another approach is to place a prior on \( K^* \), leading to the mixture-of-finite-mixtures (MFM) version of the SBM proposed by \citet{geng2019probabilistic}. Unlike the MFM, the infinite mixture model assumes \( K = \infty \), meaning that an increasing number of nodes would result in an unbounded number of groups. 

Interestingly, the DM prior, the DP prior, and the MFM prior are all examples of Gibbs-type priors (discussed below), which are notable for their analytical and computational tractability. See \citet{deblasi2015gibbs} for a comprehensive overview and \citet{legramanti2022extended} for a concise introduction.

\subsubsection{Extended stochastic block models}\label{sec_ESBM}

Despite their foundational role in network analysis, SBMs often fail to fully capture the structural complexity of real-world networks. These models assume that nodes belong to predefined, mutually exclusive communities, with connection probabilities determined solely by group membership. However, many networks exhibit more intricate architectures, including core-periphery structures, disassortative patterns, and overlapping or hierarchical communities. \citet{legramanti2022extended} introduce a more unified and adaptable framework known as extended stochastic block models (ESBMs) to address these limitations. This approach incorporates Gibbs-type priors on the partitioning process, allowing the number of groups to be finite, random, or infinite rather than fixed in advance \citep{gnedin2010species}.

ESBMs incorporate node attributes through product partition models \citep{hartigan1990partition}, allowing for the structured inclusion of external covariates such as individual roles, affiliations, or geographic locations. This feature is particularly valuable in applications like criminal network analysis, where uncovering hidden organizational hierarchies requires models that account for both structural and contextual information \citep{legramanti2022extended}. Additionally, ESBMs leverage Bayesian nonparametrics and collapsed Gibbs sampling to improve estimation accuracy, uncertainty quantification, and model-based inference. By integrating hierarchical extensions \citep{peixoto2014hierarchical} and accommodating degree heterogeneity \citep{karrer2011stochastic}, ESBMs offer a powerful and versatile framework for analyzing complex networks, effectively addressing many of the limitations of traditional SBMs.

Gibbs-type prior distributions are defined on the space of unlabeled community indicators \( \boldsymbol{\xi} \), which represents the set of all partitions of nodes into any number of groups, regardless of how the groups are labeled. A probability distribution function \( p(\boldsymbol{\xi}) \) is said to be of Gibbs-type if and only if  
\[
p(\boldsymbol{\xi}) = \textsf{w}_{n,K} \prod_{k=1}^{K} (1 - \sigma)^{n_k - 1},
\]
where \( n_k \) is the number of nodes in cluster \( k \), \( \sigma < 1 \) is the discount parameter, and  \( \{ \textsf{w}_{n,K} : 1 \leq K \leq n \} \) is a collection of nonnegative weights satisfying the recursion  
\[
\textsf{w}_{n,K} = (n - K\sigma) \textsf{w}_{n+1,K} + \textsf{w}_{n+1,K+1}, \quad \text{with } \textsf{w}_{1,1} = 1.
\]
Additionally, the ascending factorial is defined as $(a)_m = a(a + 1) \cdots (a + m - 1)$, for $a > 0$ and $m \geq 1$, with the convention \( (a)_0 = 1 \). The class of random partitions induced by Gibbs-type priors coincides with exchangeable product partition models \citep{lijoi2007controlling}.

Gibbs-type priors \citep{legramanti2022extended} form a broad yet tractable class, whose predictive distribution enables a sequential and interpretable allocation of membership indicators:  
\[
\Pr(\xi_{n+1} = k \mid \boldsymbol{\xi}) \propto  
\begin{cases}  
\textsf{w}_{n+1,K} (n_k - \sigma), & k = 1, \dots, K, \\  
\textsf{w}_{n+1,K+1}, & k = K+1,  
\end{cases}  
\]  
where $K$ is the number of non-empty communities. This formulation provides an intuitive interpretation of the group assignment process as a simple seating mechanism, where a new node joins an existing cluster with probability proportional to its size \( n_k \), adjusted by \( \sigma \) and rescaled by \( \textsf{w}_{n+1,K} \). Alternatively, it forms a new cluster with probability proportional to \( \textsf{w}_{n+1,K+1} \).

As shown in \cite{legramanti2022extended}, the following cases can be obtained as special cases of the predictive distribution given above:
\begin{itemize}

    \item \textbf{Dirichlet-Multinomial} (DM) prior: Given that \( \boldsymbol{\xi} \) follows a DM prior, the cluster assignment probability for a new observation \( \xi_{n+1} \) is  
    \[
    \Pr(\xi_{n+1} = k \mid \boldsymbol{\xi}) =
    \begin{cases}  
    \frac{n_k + \frac{\alpha}{K}}{n + \alpha}, & \text{for } k = 1, \dots, K^*, \\  
    \frac{\frac{\alpha}{K}}{n + \alpha}, & \text{for } k = K^*+1.
    \end{cases}  
    \]
    where \( n_k \) is the number of previous assignments to cluster \( k \), \( K \) is the total number of predefined clusters (a fixed upper bound), \( K^* \) is the number of currently occupied clusters, and \( \alpha \) is the concentration parameter of the Dirichlet prior.  
    
    Unlike nonparametric models, where new observations can introduce an unlimited number of clusters, in the DM prior, assigning an observation to a ``new cluster'' means activating one of the previously unoccupied clusters within the fixed \( K \) possibilities. The probability of such an assignment is proportional to the prior concentration \( \alpha \) and the number of currently unoccupied clusters, ensuring that the total number of occupied clusters \( K^* \) never exceeds \( K \).
    
    \item \textbf{Dirichlet process} (DP) prior: Given that \( \boldsymbol{\xi} \) follows a DP prior, the cluster assignment probability for a new observation \( \xi_{n+1} \) follows the Chinese restaurant process (CRP) representation, given by  
    \[
    \Pr(\xi_{n+1} = k \mid \boldsymbol{\xi}) =  
    \begin{cases}  
    \frac{n_k}{n + \alpha}, & \text{for } k = 1, \dots, K^*, \\  
    \frac{\alpha}{n + \alpha}, & \text{for } k = K^*+1,
    \end{cases}  
    \]
    where \( n_k \) is the number of previous assignments to cluster \( k \), \( K^* \) is the number of currently occupied clusters, \( n \) is the total number of observations assigned so far, and \( \alpha \) is the concentration parameter of the Dirichlet process.  

    This formulation derives from the CRP metaphor, where customers (observations) enter a restaurant with an unbounded number of tables (clusters). A new customer joins an existing table (i.e., an already occupied cluster) with probability proportional to the number of customers already seated there (\( n_k \)), reinforcing a rich-get-richer dynamic. Alternatively, the customer chooses a new table with probability proportional to \( \alpha \), introducing a new cluster.  

    Unlike finite mixture models, where the number of clusters is predetermined or bounded, in the DP prior, the number of clusters grows dynamically with the data, allowing for a potentially infinite number of groups. The probability of forming a new cluster decreases as more observations are added, ensuring that the expected number of occupied clusters grows approximately as \( \alpha \log(n) \). This nonparametric structure allows DP-based models to adaptively infer the number of clusters.

    \item \textbf{Pitman-Yor process} (PYP) prior: Given that \( \boldsymbol{\xi} \) follows a PYP prior, the cluster assignment probability for a new observation \( \xi_{n+1} \) follows the two-parameter CRP representation, given by  
    \[
    \Pr(\xi_{n+1} = k \mid \boldsymbol{\xi}) =  
    \begin{cases}  
    \frac{n_k - \sigma}{n + \alpha}, & \text{for } k = 1, \dots, K^*, \\  
    \frac{K^* \sigma + \alpha}{n + \alpha}, & \text{for } k = K^*+1,
    \end{cases}  
    \]
    where \( n_k \) is the number of previous assignments to cluster \( k \), \( K^* \) is the number of currently occupied clusters, \( n \) is the total number of observations assigned so far, \( \alpha \) is the concentration parameter, and \( \sigma \) is the discount parameter, satisfying \( 0 \leq \sigma < 1 \).  

    This formulation generalizes the CRP, introducing a discount parameter \( \sigma \) that controls the reinforcement effect. A new customer (observation) joins an existing table (cluster) with probability proportional to \( n_k - \sigma \), reducing the rich-get-richer effect compared to the standard CRP. Alternatively, the customer chooses a new table with probability proportional to \( K^* \sigma + \alpha \), making the formation of new clusters more flexible.  

    The marginal distribution of \( \boldsymbol{\xi} \) under the PYP prior follows a generalization of the Ewens sampling formula, incorporating the discount parameter \( \sigma \). Specifically, the probability of a partition \( \boldsymbol{\xi} \) forming \( K^* \) distinct clusters with cluster sizes \( (n_1, \dots, n_{K^*}) \) is given by:
    \[
    \Pr(\boldsymbol{\xi}) = \frac{1}{\Gamma(n + \alpha)}\prod_{k=1}^{K^*} (\alpha + \sigma(k-1)) \prod_{k=1}^{K^*} \Gamma(n_k - \sigma).
    \]
    This formulation differs from the standard DP prior, where the discount parameter is \( \sigma = 0 \). The introduction of \( \sigma \) alters the cluster size distribution, leading to a greater tendency for smaller clusters and a heavier tail, reflecting the power-law behavior inherent in the PYP prior.

    Unlike the DP, which leads to an expected number of clusters growing as \( \alpha \log(n) \), the PYP exhibits power-law behavior, where the number of clusters scales approximately as \( \alpha n^\sigma \). This distinction is particularly important in applications where rare but significant categories are frequently encountered, as it enables a more flexible partitioning of observations and better captures the underlying structure.  
    
    \item \textbf{Gnedin process} (GNP) prior: Given that \( \boldsymbol{\xi} \) follows a GNP prior, the cluster assignment probability for a new observation \( \xi_{n+1} \) follows the Gnedin CRP representation, given by  
    \[
    \Pr(\xi_{n+1} = k \mid \boldsymbol{\xi}) =  
    \begin{cases}  
    \frac{(n_k + 1)(n - K^* + \gamma)}{n(n+\gamma)}, & \text{for } k = 1, \dots, K^*, \\  
    \frac{K^*(K^* - \gamma)}{n(n+\gamma)}, & \text{for } k = K^*+1,
    \end{cases}  
    \]
    where \( n_k \) is the number of previous assignments to cluster \( k \), \( K^* \) is the number of currently occupied clusters, \( n \) is the total number of observations assigned so far, and \( \gamma \) is a parameter controlling the probability of forming new clusters, satisfying \( 0 < \gamma < 1 \).  

    Unlike the DP and the PYP, which allow the number of clusters to grow indefinitely, the GNP introduces a self-regulating mechanism that limits the total number of clusters even as \( n \to \infty \). This is achieved through the term \( (K^* - \gamma) \), which gradually reduces the probability of creating new clusters as the number of occupied clusters increases, effectively stabilizing the number of groups over time.  

    The marginal distribution of \( \boldsymbol{\xi} \) under the GNP follows a refinement of the Ewens sampling formula, given by:
    \[
    \Pr(\boldsymbol{\xi}) = \sum_{k=1}^{\infty} \Pr(K^* = k) \, p_{\textsf{DM}}(\boldsymbol{\xi} \mid \alpha = 1, K = k),
    \]
    where $\Pr(K^* = k) = \gamma(1 - \gamma)_{k-1}/k!$, and \( p_{\textsf{DM}}(\boldsymbol{\xi} \mid \alpha = 1, K = k) \) is the DM distribution with parameters \( \alpha = 1 \) and \( K = k \). This formulation highlights a key feature of the GNP: It can be seen as a randomized version of the DM model, where the number of occupied clusters \( K^* \) is not fixed but follows a specific heavy-tailed prior distribution.  

    Unlike models such as the DP and PYP, which always assume an infinite number of potential clusters, the GNP explicitly quantifies uncertainty in the total number of groups. In fact, the expected number of clusters under the GNP prior satisfies $\textsf{E}(K^*) = \sum_{k=1}^{n} k \Pr(K^* = k)$, which ensures a more parsimonious representation of partitions while preserving robustness to model complexity. This self-regulating growth makes the GNP particularly suitable for applications where the total number of clusters is expected to remain finite yet unknown. 
    
\end{itemize}

\subsection{Latent space models}

Another widely used approach to network modeling is the latent space model, introduced by \citet{hoff2002latent} and further developed by \citet{handcock2007modeling} and \citet{krivitsky2009representing}. In this framework, each node is assigned a latent position in a Euclidean space, and the probability of an edge forming between two nodes depends on their proximity in this latent space. Extensions for community detection emerge naturally by applying mixture models to cluster nodes with similar latent positions \citep{handcock2007modeling}.

Specifically, latent space models leverage the use of random effects in generalized linear models to capture complex network structures. For conditionally independent $y_{i,j}$, the interaction probabilities are:
\[
\textsf{Pr}(y_{i,j}=1\mid \boldsymbol{\beta}, \gamma_{i,j}, \boldsymbol{x}_{i,j} ) = \Phi(\boldsymbol{x}^\top_{i,j}\boldsymbol{\beta} + \gamma_{i,j}),\qquad 1\leq i<j \leq n,
\]  
where $\boldsymbol{\beta}=(\beta_1,\ldots,\beta_P)$ represents fixed effects, $\boldsymbol{x}^\top_{i,j}\boldsymbol{\beta}$ captures patterns from covariates $\boldsymbol{x}_{i,j}$, and $\gamma_{i,j}$ accounts for unobserved effects. As noted in \citet{hoff-2008}, a jointly exchangeable random effects matrix $[\gamma_{i,j}]$ can be expressed as $\gamma_{i,j} = \alpha(\boldsymbol{u}_i, \boldsymbol{u}_j)$, where $\boldsymbol{u}_1, \ldots, \boldsymbol{u}_n$ are independent latent variables. The function $\alpha(\cdot,\cdot)$ is central to modeling relational data. Several formulations for $\alpha(\cdot,\cdot)$ have been proposed in the literature. Here, we focus on the distance and bilinear formulations. For a comprehensive review of latent space modeling, see \citet{sosa2021review}.

\subsubsection{Distance models}

\citet{hoff2002latent} propose that each actor \(i\) occupies a position \(\boldsymbol{u}_i \in \mathbb{R}^K\) in a Euclidean space, where the probability of an edge increases as actors become closer. This is modeled as \(\alpha(\boldsymbol{u}_i, \boldsymbol{u}_j) = -\|\boldsymbol{u}_i - \boldsymbol{u}_j\|\), where \(\|\cdot\|\) denotes the Euclidean norm. Such distance-based structures naturally induce homophily (stronger ties between nodes with similar characteristics) and enable effective visualization of social networks. However, these models may be less suitable for networks with high clustering. In its simplest form, without covariates, the distance model is given by:
\[
y_{i,j} \mid \zeta, \boldsymbol{u}_i, \boldsymbol{u}_j \overset{\text{ind}}{\sim} \textsf{Ber}\left(\Phi\left(\zeta - \|\boldsymbol{u}_i - \boldsymbol{u}_j\|\right)\right),
\]  
where \(\zeta\) represents the average edge propensity, and \(\boldsymbol{u}_1, \dots, \boldsymbol{u}_I\) are latent positions in \(\mathbb{R}^K\). The priors are specified as \(\zeta \mid \tau^2 \sim \textsf{N}(0, \tau^2)\) and \(\boldsymbol{u}_i \mid \sigma^2 \sim \textsf{N}(\boldsymbol{0}, \sigma^2 \mathbf{I})\), where \(\mathbf{I}\) is the identity matrix, and finally, the hyperpriors are given by \(\tau^2 \sim \textsf{IG}(a_\tau, b_\tau)\) and \(\sigma^2 \sim \textsf{IG}(a_\sigma, b_\sigma)\). This hierarchical specification introduces flexibility in modeling network connectivity and latent position uncertainty. Numerous studies have explored distance models and their extensions, including the works of \cite{handcock2007model}, \cite{krivitsky2009representing}, \cite{ciminelli2019social}, and \cite{kaur2023latent}.

\subsubsection{Bilinear models}

\cite{hoff2005bilinear}, building on the latent space framework underlying distance models, proposes that interaction probabilities are governed by symmetric multiplicative effects. In this formulation, the interaction between actors \(i\) and \(j\) is represented by the inner product of their latent position vectors, given by \(\alpha(\boldsymbol{u}_i, \boldsymbol{u}_j) = \boldsymbol{u}_i^\top \boldsymbol{u}_j\). As noted by \cite{hoff-2008}, bilinear models generalize distance-based models (though not class-based models) and offer greater flexibility in capturing structural features such as varying levels of balance and clusterability. In the bilinear model, the sampling distribution is given by
\[
y_{i,j} \mid \zeta, \boldsymbol{u}_i, \boldsymbol{u}_j \overset{\text{ind}}{\sim} \textsf{Ber}\left( \Phi\left(\zeta + \boldsymbol{u}_i^\top \boldsymbol{u}_j \right) \right),
\]
where \(\zeta\) represents the average edge propensity, and \(\boldsymbol{u}_1, \dots, \boldsymbol{u}_I\) are latent positions in \(\mathbb{R}^K\). The term \(\boldsymbol{u}_i^\top \boldsymbol{u}_j = \sum_{k=1}^K u_{i,k} u_{j,k}\) captures the degree of alignment between the actors' latent characteristics: a higher inner product implies greater similarity or compatibility, thus increasing the probability of a tie. Following the prior formulation used in distance models, we assume \(\zeta \mid \tau^2 \sim \textsf{N}(0, \tau^2)\) and \(\boldsymbol{u}_i \mid \sigma^2 \overset{\text{iid}}{\sim} \textsf{N}(\boldsymbol{0}, \sigma^2 \mathbf{I})\), with \(\tau^2 \sim \textsf{IG}(a_\tau, b_\tau)\) and \(\sigma^2 \sim \textsf{IG}(a_\sigma, b_\sigma)\). Several extensions and refinements of distance and bilinear models have been proposed in the literature, including \cite{hoff2009multiplicative}, \cite{ward2011network}, \cite{minhas2019inferential}, and \cite{hoff2021additive}.

Finally, \citet{hoff-2008} and \citet{hoff2009multiplicative} model relationships between nodes using a weighted inner product of latent vectors \(\boldsymbol{u}_i \in \mathbb{R}^K\), defining the interaction term as \(\alpha(\boldsymbol{u}_i, \boldsymbol{u}_j) = \boldsymbol{u}_i^\top \mathbf{\Lambda} \boldsymbol{u}_j\), where \(\mathbf{\Lambda}\) is a \(K \times K\) diagonal matrix. Known as the eigen model, this approach extends latent class and distance models by efficiently capturing key network features and accommodating both positive and negative forms of homophily. Although the parameters \(\lambda_1, \ldots, \lambda_K\) conveniently control the contribution of each latent dimension to the overall structure, this flexibility comes at the cost of introducing an additional set of parameters, which can complicate both the computation and interpretation.

\section{Hybrid class models} \label{sec:hybrid}

We introduce a hybrid modeling framework that extends the stochastic block model by embedding clusters into a latent Euclidean space, combining the interpretability of block models with the geometric flexibility of latent space approaches. This formulation enables parsimonious modeling of inter-group connectivity, enhances both clustering and predictive performance, and supports fully Bayesian inference via MCMC algorithms under parametric and nonparametric clustering priors. Importantly, our approach differs substantially from traditional position latent cluster models (e.g., \citealt{krivitsky2009representing}). While those models embed individual actors and impose a finite mixture prior over their latent positions to induce clustering, we reparameterize the community interaction structure directly by embedding clusters into a latent cluster-level social space. This shift better aligns with applications where group-level interactions are the primary focus.

\subsection{Modeling approach}

Our proposal is guided by two main motivations. First, we aim to represent clusters within a low-dimensional Euclidean \textit{cluster space} using group-specific latent features for modeling group-level social dynamics. Second, we seek to improve model performance in both clustering and predictive tasks by explicitly capturing relational patterns at the group level. Rather than relying only on unstructured inter- and intra-group interaction probabilities, this approach provides an integrated framework for representing network connectivity.

Under the standard SBM setup described in Section \ref{sec_stochastic_blockmodels}, we introduce additional structure into the model by reparameterizing the community parameters \(\eta_{k,\ell}\), leveraging the principles of latent space distance modeling. To achieve this, we assume that edge formation between any two actors is governed by a global propensity, which is modulated by dissimilarities arising from a distinct set of latent features that characterize intra- and inter-cluster social dynamics. Specifically, rather than embedding individual actors, we embed clusters in a low-dimensional Euclidean \textit{cluster space}, allowing the community-level interaction parameters to be expressed as  
\[
\eta_{k,\ell} = \eta - \|\boldsymbol{u}_k - \boldsymbol{u}_\ell\|, \qquad 1 \leq k < \ell \leq K,
\]
where \(K\) denotes either the number of active clusters under a DM prior or the number of non-empty clusters under nonparametric priors. Each position $\boldsymbol{u}_k = (u_{k,1}, \ldots, u_{k,Q})$, for $k = 1,\ldots,K$, is embedded in a Euclidean space of dimension $Q$, known as the latent dimension, which plays a crucial role in latent space modeling. We call this model, the \textit{class-distance model} (CDM).

Here, the fixed effect $\eta$ captures the overall propensity for connectivity across the network, serving as the baseline for edge formation when the distance between clusters is zero. The cluster-specific latent positions $\boldsymbol{u}_1, \dots, \boldsymbol{u}_K$ define a geometric embedding of the clusters, where the Euclidean distance $\|\boldsymbol{u}_k - \boldsymbol{u}_\ell\|$ measures their dissimilarity. Actors belonging to clusters that are closer in this space have a higher probability of interaction, while those in more distant clusters are less likely to connect. This formulation provides a coherent representation of both intra- and inter-cluster connectivity, allowing the model to capture local clustering behavior as well as broader structural patterns in the network.

This specification explicitly incorporates clustering effects into the network structure while introducing a continuous notion of community distance. By modeling distances between clusters rather than individuals, the approach substantially reduces the number of parameters compared to standard latent distance models, which assign a latent position to each actor. Assigning latent components at the cluster level enhances computational efficiency, as the number of clusters is typically much smaller than the number of actors in the network. Moreover, this formulation preserves the core structure of a stochastic block model (SBM), ensuring that community detection remains central, while simultaneously leveraging latent space modeling to provide a richer characterization of both intra- and inter-cluster social dynamics. By integrating these elements, the model effectively balances complexity reduction with structural flexibility.

It is important to note that this strategy does not support the direct visualization of clusters for three main reasons. First, the latent positions are not identifiable. This issue is typically addressed by concentrating inference on a specific configuration of latent positions, obtained through a Procrustes transformation that aligns each sample to a fixed reference configuration of the latent space \citep{hoff2002latent}. Second, the cluster assignments are not identifiable. This is a well-known issue in the clustering literature, since the same partition can be represented using different arrangements of cluster indicators. Third, and related to the previous point, because latent positions are assigned to clusters rather than individuals, and clusters can appear or disappear during the training algorithm, the number of latent positions varies across iterations. This variation prevents the construction of a consistent geometric representation.

Now, to perform fully Bayesian inference on the model parameters, it is necessary to specify prior distributions for the random effects and latent positions. Specifically, the global connectivity parameter \(\eta\) is assigned a normal prior \(\eta \mid \tau^2 \sim \textsf{N}(0, \tau^2)\), where \(\tau^2\) governs the variability in overall network connectivity. The latent positions \(\boldsymbol{u}_k\), which define the geometric embedding of clusters in the social space, follow a prior distribution \(\boldsymbol{u}_k \mid \sigma^2 \overset{\text{iid}}{\sim} \textsf{N}(\mathbf{0}, \sigma^2 \mathbf{I})\), where \(\sigma^2\) controls the spread of clusters in the latent space. The hyperpriors are specified as \(\tau^2 \sim \textsf{IG}(a_\tau, b_\tau)\) and \(\sigma^2 \sim \textsf{IG}(a_\sigma, b_\sigma)\), allowing for adaptive uncertainty in the variance components, where \(a_\tau\), \(b_\tau\), \(a_\sigma\), and \(b_\sigma\) are model hyperparameters. Finally, for the cluster assignments, we consider both parametric and nonparametric prior specifications as described in Section \ref{sec_ESBM}, with hyperparameters determined on a case-specific basis.

The class-distance model can be naturally extended to a bilinear specification, giving rise to what we refer to as the \textit{class-bilinear model} (CBM), following the framework introduced by \citet{hoff2005bilinear}. In the original class-distance formulation, intra-cluster interactions are constrained by the fixed effect $\eta$, since the Euclidean distance between any cluster and itself is zero, i.e., $\|\boldsymbol{u}_k - \boldsymbol{u}_k\| = 0$. This implies that all within-cluster interaction probabilities are equal to $\Phi(\eta)$. As a result, the model lacks flexibility in differentiating intra-cluster connectivity patterns. The class-bilinear model addresses this limitation by defining the interaction between clusters $k$ and $\ell$ through a symmetric inner product of their latent vectors, given by $\eta_{k,\ell} = \eta + \boldsymbol{u}_k^\top \boldsymbol{u}_\ell$. This formulation allows intra-cluster probabilities to vary freely, since $\boldsymbol{u}_k^\top \boldsymbol{u}_k$ does not vanish and can differ across clusters. Each cluster remains embedded in a low-dimensional Euclidean space, but interactions are now governed by alignment rather than proximity. This extension preserves the computational advantages of modeling at the cluster level while expanding the model's capacity to capture complex relational patterns, including transitivity, structural balance, and overlapping group structure. It also aligns with the eigenmodel framework of \citet{hoff-2008} and \citet{hoff2009multiplicative}, in which a diagonal matrix $\mathbf{\Lambda}$ can further modulate the contribution of each latent dimension, thereby enhancing the model’s ability to reflect both positive and negative forms of homophily across communities.

\subsection{Computation}

Let $\mathbf{\Theta}$ denote the set of model parameters. The posterior distribution $p(\mathbf{\Theta} \mid \mathbf{Y})$ can be explored using Markov chain Monte Carlo (MCMC) methods (e.g., \citealt{gamerman2006markov}). The computational algorithm combines Gibbs sampling and Metropolis steps, with specific updates depending on the chosen model structure (class, class distance, or class bilinear) and the prior specification for cluster assignments (DM, DP, PYP, or GNP). In what follows, we provide a detailed description of the MCMC algorithms used to fit each model under any prior configuration.

\subsubsection{Class models}\label{sec:mcmc_class}

Consider an $I \times I$ adjacency matrix $\mathbf{Y} = [y_{i,j}]$ associated with an undirected binary network. As described in Section \ref{sec_stochastic_blockmodels}, under the standard class model we have that the likelihood of the data given the cluster assignments $\boldsymbol{\xi} = (\xi_1, \ldots, \xi_I)$ and the community-level interaction parameters $\boldsymbol{\eta} = (\eta_{1,1}, \eta_{1,2}, \ldots, \eta_{K,K})$ is given by the expression in \eqref{eq_class_likelihood}. Thus, the joint posterior distribution of the model parameters $\boldsymbol{\Theta}$ given the data $\mathbf{Y}$ is proportional to
\begin{align*}
p(\boldsymbol{\Theta} \mid \mathbf{Y}) &\propto 
\prod_{k=1}^{K} \prod_{\ell=k}^{K} \left[\Phi(\eta_{k,\ell})\right]^{s_{k,\ell}} \left[1 - \Phi(\eta_{k,\ell})\right]^{n_{k,\ell} - s_{k,\ell}} \times \prod_{k=1}^K \prod_{\ell=k}^K \textsf{N}(\eta_{k,\ell} \mid \zeta, \tau^2) \\ 
&\quad \times \textsf{N}(\zeta \mid \mu_\zeta, \sigma^2_\zeta) \times \textsf{IG}(\tau^2 \mid a_\tau, b_\tau) \times p(\boldsymbol{\xi}),
\end{align*}
where $p(\boldsymbol{\xi})$ corresponds to one of the four prior specifications for the cluster assignments (DM, DP, PYP, or GNP). The hyperparameters $\mu_\zeta$, $\sigma^2_\zeta$, $a_\tau$, and $b_\tau$ are fixed and known. Note that $p(\boldsymbol{\xi})$ involves additional hyperparameters depending on the chosen prior.

The algorithm proceeds by generating a new state $\boldsymbol{\Theta}^{(b+1)}$ from the current state $\boldsymbol{\Theta}^{(b)}$, for $b = 1, \ldots, B$, by sequentially sampling from the following full conditional distributions, each conditioned on the most recent updates of the remaining parameters:
\begin{enumerate}
\item Sample $\eta_{k,\ell}^{(b+1)}$, for $k,\ell = 1, \ldots, K$ (or \(K^*\) for nonparametric priors), using a Metropolis step, based on the full conditional distribution:
$$
\log p(\eta_{k,\ell} \mid -) \propto
s_{k,\ell} \log\left( \Phi(\eta_{k,\ell}) \right) + (n_{k,\ell} - s_{k,\ell}) \log\left( 1 - \Phi(\eta_{k,\ell}) \right) - \frac{1}{2\tau^2} (\eta_{k,\ell} - \zeta)^2.
$$
	
\item Sample $\zeta^{(b+1)}$ from $\textsf{N}(m,v^2)$, where
$$
v^2 = \left(\frac{1}{\sigma^2_\zeta} + \frac{1}{\tau^2}\,\frac{K(K+1)}{2}\right)^{-1} 
\qquad\text{and}\qquad 
m =	v^2\,\left( \frac{\mu_\zeta}{\sigma^2_\zeta} + \frac{1}{\tau^2}\sum_{k=1}^K\sum_{\ell=k}^K\eta_{k,\ell} \right)\,.
$$
	
\item Sample $(\tau^2)^{(b+1)}$ from $\textsf{IG}(c,d)$, where
$$
c = a_\tau + \frac12\, \frac{K(K+1)}2
\qquad\text{and}\qquad
d = b_\tau + \frac12\sum_{k=1}^K\sum_{\ell=k}^K (\eta_{k,\ell}-\zeta)^2.
$$

\item Sample $\xi_i$, for $i = 1, \ldots, I$, from a categorical distribution determined by the selected clustering prior (DM, DP, PYP, or GNP). We distinguish between $K$, the fixed number of clusters used under parametric priors (e.g., DM), and $K^*$, the current number of non-empty clusters in the nonparametric setting (e.g., DP, PYP, or GNP).

\begin{itemize}
    \item \textbf{Dirichlet-Multinomial} (DM) prior:

    \begin{enumerate}[label=\roman*.]
    \item Sample \(\xi_i^{(b+1)}\), for \(i = 1, \ldots, I\), from a categorical distribution over \(\{1, \ldots, K\}\) with probabilities proportional to
    \begin{align*}
    \Pr(\xi_i = k \mid -) 
    &\propto 
    \prod_{j = i+1}^{I} \Phi(\eta_{\phi(k, \xi_j)})^{y_{i,j}} \left[1 - \Phi(\eta_{\phi(k, \xi_j)})\right]^{1 - y_{i,j}} \\ 
    &\,\,\,\times \prod_{j = 1}^{i-1} \Phi(\eta_{\phi(\xi_j, k)})^{y_{j,i}} \left[1 - \Phi(\eta_{\phi(\xi_j, k)})\right]^{1 - y_{j,i}} \times \omega_k.
    \end{align*}

    \item Sample \(\boldsymbol{\omega}^{(b+1)}\) from the full conditional distribution
    \[
    p(\boldsymbol{\omega} \mid \text{rest}) = \textsf{Dir}\left( \frac{\alpha}{K} + n_1, \ldots, \frac{\alpha}{K} + n_K \right),
    \]
    where \(n_k = \sum_{i=1}^I 1(\xi_i = k)\) is the number of actors currently assigned to cluster \(k\).
    \end{enumerate}

    \item \textbf{Dirichlet process} (DP) prior:

    \begin{enumerate}[label=\roman*.]
    \item Sample \(\xi_i^{(b+1)}\), for \(i = 1, \ldots, I\), from a categorical distribution over \(\{1, \ldots, K^*, K^*+1\}\), where \(K^*\) is the current number of non-empty clusters. The probabilities are proportional to:
    \begin{align*}
    \Pr(\xi_i = k \mid -) 
    &\propto 
    \prod_{j = i+1}^{I} \Phi(\eta_{\phi(k, \xi_j)})^{y_{i,j}} \left[1 - \Phi(\eta_{\phi(k, \xi_j)})\right]^{1 - y_{i,j}} \\ 
    &\,\,\,\times \prod_{j = 1}^{i-1} \Phi(\eta_{\phi(\xi_j, k)})^{y_{j,i}} \left[1 - \Phi(\eta_{\phi(\xi_j, k)})\right]^{1 - y_{j,i}} \\ &\,\,\,\times \, \Pr(\boldsymbol{\xi}_i = k\mid\boldsymbol{\xi}_{-i}),
    \end{align*}
    for existing clusters $k = 1, \ldots, K^*$, and
    \begin{align*}
    \Pr(\xi_i = k \mid -) 
    &\propto 
    \prod_{j = i+1}^{I} \Phi(\eta_{\phi(k, \xi_j)})^{y_{i,j}} \left[1 - \Phi(\eta_{\phi(k, \xi_j)})\right]^{1 - y_{i,j}} \\ 
    &\,\,\,\times \prod_{j = 1}^{i-1} \Phi(\eta_{\phi(\xi_j, k)})^{y_{j,i}} \left[1 - \Phi(\eta_{\phi(\xi_j, k)})\right]^{1 - y_{j,i}} \\ 
    &\,\,\,\times \prod_{\ell=1}^{K} \textsf{N}(\eta^*_{k,\ell} \mid \mu, \tau^2) \\
    &\,\,\,\times \Pr(\boldsymbol{\xi}_i = k\mid\boldsymbol{\xi}_{-i}),
    \end{align*}
    for a new cluster $k = K^*+1$, where 
    \[
    \Pr(\xi_i = k \mid \boldsymbol{\xi}_{-i}) \propto 
    \begin{cases}
    n_k^{(-i)} , & \text{for } k = 1, \ldots, K^*, \\
    \alpha,     & \text{for } k = K^* + 1,
    \end{cases}
    \]
    $\alpha \in (0,\infty)$ is the concentration parameter that controls the overall tendency to form new clusters, \(n_k^{(-i)}\) is the number of actors in cluster \(k\) excluding actor \(i\), and \(\eta^*_{k,\ell}\) are interaction parameters sampled from the prior distribution \(\textsf{N}(\mu, \sigma^2)\).

    \item Update the interaction parameters \(\boldsymbol{\eta}\) by augmenting or reducing the dimension of the array according to whether a new cluster was created or an existing cluster became empty. All \(\eta_{k,\ell}\) are retained for active clusters only.

    \item Relabel the cluster assignments \(\boldsymbol{\xi}\) and reshape \(\boldsymbol{\eta}\) to ensure that cluster labels remain consecutive and that no empty clusters are retained.
    \end{enumerate}

    \item \textbf{Pitman-Yor process} (PYP) prior: The sampler follows the same structure as in the DP prior. However, under the PYP prior, the prior probability of assigning actor $i$ to cluster $k$ is proportional to
    $$
    \Pr(\xi_i = k \mid \boldsymbol{\xi}_{-i}) \propto 
    \begin{cases}
    n_k^{(-i)} - \sigma, & \text{for } k = 1, \ldots, K^*, \\
    \sigma K^* + \alpha, & \text{for } k = K^* + 1,
    \end{cases}
    $$
    where $\sigma \in [0,1)$ is the discount parameter, $n_k^{(-i)}$ is the number of actors in cluster $k$ excluding actor $i$, and $K^*$ is the current number of non-empty clusters.

    \item \textbf{Gnedin process} (GNP) prior: The sampler follows the same structure as in the DP and PYP priors. However, under the GNP prior, the prior probability of assigning actor $i$ to cluster $k$ is proportional to
    $$
    \Pr(\xi_i = k \mid \boldsymbol{\xi}_{-i}) \propto 
    \begin{cases}
    (n_k^{(-i)} + 1)(I - K^* + \gamma), & \text{for } k = 1, \ldots, K^*, \\[4pt]
    K^*(K^* - \gamma), & \text{for } k = K^* + 1,
    \end{cases}
    $$
    where $\gamma \in (0,1)$ is a discount parameter controlling cluster reinforcement and the formation of new clusters, $n_k^{(-i)}$ is the number of actors currently assigned to cluster $k$, excluding actor $i$, and $K^*$ is the number of non-empty clusters at the current iteration.
\end{itemize}
\end{enumerate}

The hyperparameters of the clustering priors, namely \(\alpha\), \(\sigma\), and \(\gamma\), are crucial for model performance. These parameters are carefully selected, as described in Section \ref{sec:hyperparameter_elicitation}, based on the expected number of occupied clusters a priori.

\subsubsection{Hybrid class models}\label{sec:mcmc_modified_class}

For both the CDM (Class-Distance Model) and the CBM (Class-Bilinear Model), the MCMC algorithm follows the same general structure as the class model described in the previous section, but Steps 1 and 4 must be modified. This adjustment is necessary because we now introduce cluster-specific latent effects $\boldsymbol{u}_k$ embedded in a $Q$-dimensional Euclidean space, where $k$ ranges from 1 to $K$ under the DM prior and from 1 to $K^*$ under nonparametric priors, to reparameterize the interaction terms $\eta_{k,\ell}$ as $\eta_{k,\ell} = \eta + h(\boldsymbol{u}_k, \boldsymbol{u}_\ell)$. The function $h(\cdot, \cdot)$ is defined as $h(\boldsymbol{u}_k, \boldsymbol{u}_\ell) = -\|\boldsymbol{u}_k - \boldsymbol{u}_\ell\|$ for the CDM and $h(\boldsymbol{u}_k, \boldsymbol{u}_\ell) = \boldsymbol{u}_k^\top \boldsymbol{u}_\ell$ for the CBM. In addition, we must incorporate sampling steps for additional parameters that enrich the reparameterization of the interaction terms.

Under the hybrid class models, the joint posterior distribution of the model parameters $\boldsymbol{\Theta}$ given the data $\mathbf{Y}$ is proportional to
\begin{align*}
p(\boldsymbol{\Theta} \mid \mathbf{Y}) &\propto 
\prod_{k=1}^{K} \prod_{\ell=k}^{K} \left[\Phi(\eta_{k,\ell})\right]^{s_{k,\ell}} \left[1 - \Phi(\eta_{k,\ell})\right]^{n_{k,\ell} - s_{k,\ell}} \times \prod_{k=1}^K \textsf{N}(\boldsymbol{u}_k\mid\boldsymbol{0},\sigma^2\mathbf{I}) \\ 
&\quad \times \textsf{N}(\eta \mid \mu_\eta, \sigma^2_\eta) \times \textsf{IG}(\sigma^2 \mid a_\sigma, b_\sigma) \times p(\boldsymbol{\xi}),
\end{align*}
where $\eta_{k,\ell} = \eta + h(\boldsymbol{u}_k, \boldsymbol{u}_\ell)$, and $\mu_\eta$, $\sigma^2_\eta$, $a_\sigma$, and $b_\sigma$ are the model hyperparameters. In this way, the algorithm proceeds by generating a new state $\boldsymbol{\Theta}^{(b+1)}$ from the current state $\boldsymbol{\Theta}^{(b)}$, for $b = 1, \ldots, B$, by sequentially sampling from the following full conditional distributions, each conditioned on the most recent updates of the remaining parameters:
\begin{enumerate}
    \item Sample \(\boldsymbol{u}_k^{(b+1)}\) for \(k = 1, \ldots, K\) (or \(K^*\) for nonparametric priors) using a Metropolis step, based on the full conditional:
    $$
    \log p(\boldsymbol{u}_k \mid -) \propto 
    \sum_{(r,s) \in \mathcal{D}_k} \big[\, y_{r,s} \log \Phi(\eta_{r,s}) + (1 - y_{r,s}) \log (1 - \Phi(\eta_{r,s}) \, \big] - \frac{1}{2\sigma^2} \|\boldsymbol{u}_k\|^2 ,
    $$
    where \(\mathcal{D}_k = \{(r,s) : \xi_r = k \text{ or } \xi_s = k,\ r < s\}\) and $\eta_{r,d} = \eta + h(\boldsymbol{u}_{\xi_r}, \boldsymbol{u}_{\xi_s})$.
    
    \item Sample \(\eta^{(b+1)}\) using a Metropolis step, based on the full conditional distribution:
    $$
    \log p(\eta \mid -) \propto 
    \sum_{k=1}^{K} \sum_{\ell=k}^{K} \big[\, s_{k,\ell} \log \Phi(\eta_{k,\ell}) + (n_{k,\ell} - s_{k,\ell}) \log (1 - \Phi(\eta_{k,\ell})) \, \big] - \frac{1}{2\sigma^2_\eta}(\eta - \mu_\eta)^2,
    $$
    where $\eta_{k,\ell} = \eta + h(\boldsymbol{u}_k, \boldsymbol{u}_\ell)$.
    
    \item Sample \((\sigma^2)^{(b+1)}\) from $\textsf{IG}(c,d)$, where:
    \[
    c = a_\sigma + \frac{KQ}{2}
    \qquad\text{and}\qquad
    d = \ b_\sigma + \frac{1}{2} \sum_{k=1}^{K} \|\boldsymbol{u}_k\|^2.
    \]

    \item Sample \(\xi_i\), for \(i = 1, \ldots, I\), from a categorical distribution determined by the selected clustering prior (DM, DP, PYP, or GNP). The likelihood is computed using the reparameterized interaction terms. For the DM prior, also sample $\boldsymbol{\omega}^{(b+1)} \sim \textsf{Dir} \left( \frac{\alpha}{K} + n_1, \ldots, \frac{\alpha}{K} + n_K \right)$,  where \(n_k = \sum_{i=1}^I 1(\xi_i = k)\) is the number of actors currently assigned to cluster \(k\). For nonparametric priors (DP, PYP, GNP), update the number of active clusters \(K^*\), relabel \(\boldsymbol{\xi}\), and add or remove latent vectors \(\boldsymbol{u}_k\) accordingly.
\end{enumerate}

\subsection{Hyperparameter elicitation}\label{sec:hyperparameter_elicitation}

In this section, we discuss the selection of model hyperparameters and the latent dimension to ensure appropriate model performance. On the one hand, for the class model, the hyperparameters include \(\mu_\zeta\), \(\sigma^2_\zeta\), \(a_\tau\), and \(b_\tau\), which correspond to the prior mean and variance of the global connectivity parameter \(\zeta\), and the shape and scale parameters of the Inverse-Gamma prior on the variance component \(\tau^2\). For the hybrid class models, the hyperparameters are \(a_\sigma\), \(b_\sigma\), \(a_\tau\), and \(b_\tau\), which control the priors on the variance of the cluster-specific latent effects \(\boldsymbol{u}_k\) and the global connectivity parameter \(\eta\). On the other hand, regarding the clustering prior, the relevant hyperparameters vary depending on the specification: \(\alpha\) for both the DM and DP priors; \(\alpha\) and \(\sigma\) for the PYP prior; and \(\gamma\) for the GNP prior. While it is possible to assign prior distributions to these parameters, we choose to treat them as fixed hyperparameters in order to enhance computational tractability and maintain model parsimony. This approach avoids introducing additional layers of complexity in the posterior distribution, which can significantly increase the computational burden without substantial gains in inference for the clustering structure.

Regarding the class model, we set $\mu_\zeta = 0$ and $\sigma^2_\zeta = 3$ to specify a weakly informative prior on the global connectivity parameter $\zeta$, centered at zero with moderate dispersion. This specification yields a 95\% prior interval for $\zeta$ approximately within $(-3.3, 3.3)$, thereby allowing for both sparse and moderately dense network structures. Additionally, we set $a_\tau = 3$ and $b_\tau = 2$ to induce a prior on the variance parameter $\tau^2$ with finite mean and variance. This choice favors moderate variability in the interaction terms $\eta_{k,\ell}$, while preserving enough flexibility for the data to inform the extent of heterogeneity across cluster interactions.

In turn, for the hybrid class models, we set $a_\sigma = 3$ and $b_\sigma = 2$ to define a moderately informative prior on the variance $\sigma^2$ of the cluster-specific latent effects $\boldsymbol{u}_k$, encouraging moderate dispersion in the latent space while avoiding overly diffuse or implausibly concentrated configurations. Furthermore, we set $a_\tau = 3$ and $b_\tau = 2$ for the prior on the variance $\tau^2$ of the global intercept $\eta$, which governs the prior uncertainty around $\eta$, promoting regularization of the global connectivity level while preserving sufficient flexibility to adapt to diverse network structures. 

In all hybrid class-type models, we fix the latent dimension to $Q = 4$. This choice balances parsimony, computational efficiency, and the need to capture sufficient latent structure without overfitting. A larger latent dimension may lead to overly complex models with poor generalization, while smaller values may underrepresent relevant structure in the data. Alternatively, we could have fitted the model for a range of values of $Q$ and selected the optimal dimension using a model selection criterion such as the Watanabe-Akaike Information Criterion (WAIC) or a predictive measure like the Area Under the Curve (AUC) as in \cite{sosa2022latent}. A more ambitious and theoretically principled approach would be to place a prior on $Q$ and implement a transdimensional MCMC algorithm, such as reversible jump MCMC \citep{green2009reversible}. Although these alternatives are appealing and worth exploring, we opt to fix $Q$ to ensure identifiability, simplify inference, and reduce computational burden, while still providing a flexible yet tractable representation of the latent space.

Finally, to calibrate the hyperparameters of the clustering priors, we begin by specifying a target value for the expected number of non-empty clusters \(\textsf{E}[K^*]\), which we set to \(K = \lfloor I / \bar{d} \rfloor\), where \(\bar{d}\) denotes the average node degree. For the DM prior, where the number of clusters \(K\) is fixed, we compute \(\alpha\) by numerically solving the identity \(\textsf{E}[K^*] = \alpha \log((\alpha + I)/\alpha)\). For the DP prior, where the expected number of clusters grows logarithmically as \(\textsf{E}[K^*] \approx \alpha \log I\), we set \(\alpha = K^*/\log I\). Under the PYP prior, where \(\textsf{E}[K^*] \approx \alpha I^\sigma / \sigma\), we let \(\sigma = \log K^* / \log I\) and then compute \(\alpha = \sigma K^* / I^\sigma\). For the GNP prior, using the asymptotic approximation \(\textsf{E}[K^*] \approx \gamma \log I / (1 - \gamma)\), we solve for \(\gamma = K^*/(K^* + \log I)\).

\subsection{Performance evaluation}\label{sec:evaluation}

Following \cite{legramanti2022extended}, we evaluate model performance using a comprehensive set of metrics that assess clustering accuracy, uncertainty quantification, and predictive ability. A central criterion is the Variation of Information (VI;\citealt{meilua2007comparing} and \citealt{wade2018bayesian}), which quantifies the dissimilarity between two partitions by comparing their individual and joint entropies. The VI ranges from 0 to $\log_2 I$, where $I$ is the number of nodes, and decreases as the overlap between partitions increases. We compute the posterior mean $\textsf{E}[\text{VI}(\mathbf{z}, \mathbf{z}_0) \mid \mathbf{Y}]$, where $\mathbf{z}_0$ denotes the true partition, to summarize the average posterior clustering error. In addition, we calculate $\text{VI}(\hat{\mathbf{z}}, \mathbf{z}_b)$, the distance between the estimated partition $\hat{\mathbf{z}}$ and the boundary $\mathbf{z}_b$ of the 95\% credible ball, which is defined as the smallest set of partitions within a given VI radius around $\mathbf{z}$ that accumulates at least 95\% of the posterior probability mass. This approach offers a principled alternative to conventional uncertainty quantification tools such as the posterior similarity matrix and enables direct inference on the space of partitions. We also report the posterior distribution of the number of non-empty clusters $H$, summarized by its median and interquartile range, as a measure of the model’s capacity to recover the true cluster complexity while balancing parsimony.

When ground truth clusterings are available, we compute the Adjusted Rand Index (ARI; \citealt{warrens2022understanding}) to evaluate the agreement between the estimated and true partitions. The ARI adjusts for chance agreement and ranges from 0, indicating random labeling, to 1, indicating perfect correspondence, offering an interpretable and widely used benchmark for clustering accuracy. In addition, we assess the pairwise False Discovery Rate (FDR) and False Negative Rate (FNR), derived from the co-clustering matrices of the estimated and true partitions. The FDR measures the proportion of node pairs incorrectly grouped together, while the FNR quantifies the proportion of truly co-clustered pairs that are incorrectly separated. These metrics jointly provide nuanced information about the model’s tendency to over-cluster or under-cluster the data.

We also incorporate the Widely Applicable Information Criterion (WAIC; \citealt{watanabe2013widely}) to assess out-of-sample predictive accuracy. WAIC is a fully Bayesian model selection criterion computed from the log-likelihood evaluated at posterior samples, with an adjustment for model complexity via the variance of the log-likelihood. Lower WAIC values indicate better predictive performance. In addition, we evaluate the model’s ability to reconstruct the observed network by comparing the posterior mean edge probability matrix $\mathbf{P}$ with the observed adjacency matrix $\mathbf{Y}$. This comparison is carried out using three complementary metrics: mean squared error (MSE), log-loss (or binary cross-entropy), and the area under the ROC curve (AUC). The MSE quantifies the average squared deviation between the observed edges and the predicted edge probabilities, emphasizing calibration accuracy. Log-loss evaluates the quality of the probabilistic predictions by penalizing both over- and under-confident estimates, particularly when predicted probabilities diverge from the observed binary outcomes. Finally, AUC measures the model’s ability to discriminate between connected and non-connected node pairs, based on how well the predicted probabilities rank the presence of edges. AUC is especially informative when used to assess the model’s ranking performance, with values near one indicating strong separation between true edges and non-edges. 

\section{Simulation}\label{sec:simulation}

We compare our hybrid class models to the standard class model under four different clustering priors. That is, three underlying class models: the standard class model (CM), the class-distance model (CDM), and the class-bilinear model (CBM), each evaluated under four clustering priors: the Dirichlet-Multinomial (DM), Dirichlet Process (DP), Pitman–Yor Process (PYP), and Gnedin Process (GNP). Thus, for each dataset, we fit a total of 12 models corresponding to every combination of model type and clustering prior. All models are estimated using the MCMC algorithms described in Sections \ref{sec:mcmc_class} and \ref{sec:mcmc_modified_class}, with hyperparameter specifications as outlined in Section \ref{sec:hyperparameter_elicitation}. The results presented in this section are based on $B = 10{,}000$ posterior samples, obtained by thinning every 50 iterations after a burn-in period of $100{,}000$ iterations. The Markov chains exhibit satisfactory mixing behavior and displayed the typical features of clustering models, namely strong autocorrelation within modes and occasional transitions between multiple high-probability modes. To optimize storage and computational resources, we only retain the sampled values of the interaction probabilities $\eta_{k,\ell}$ and the cluster assignments $\xi_i$. This reduced representation suffices for characterizing the posterior cluster structure and computing all evaluation metrics necessary for model comparison. All the code necessary to reproduce our results is freely available at \url{https://github.com/jstats1702/modified-class-models}.

\begin{figure}[!htb]
    \centering
    \subfigure[Scenario 1: Socio matrix.] {\includegraphics[scale=0.43]{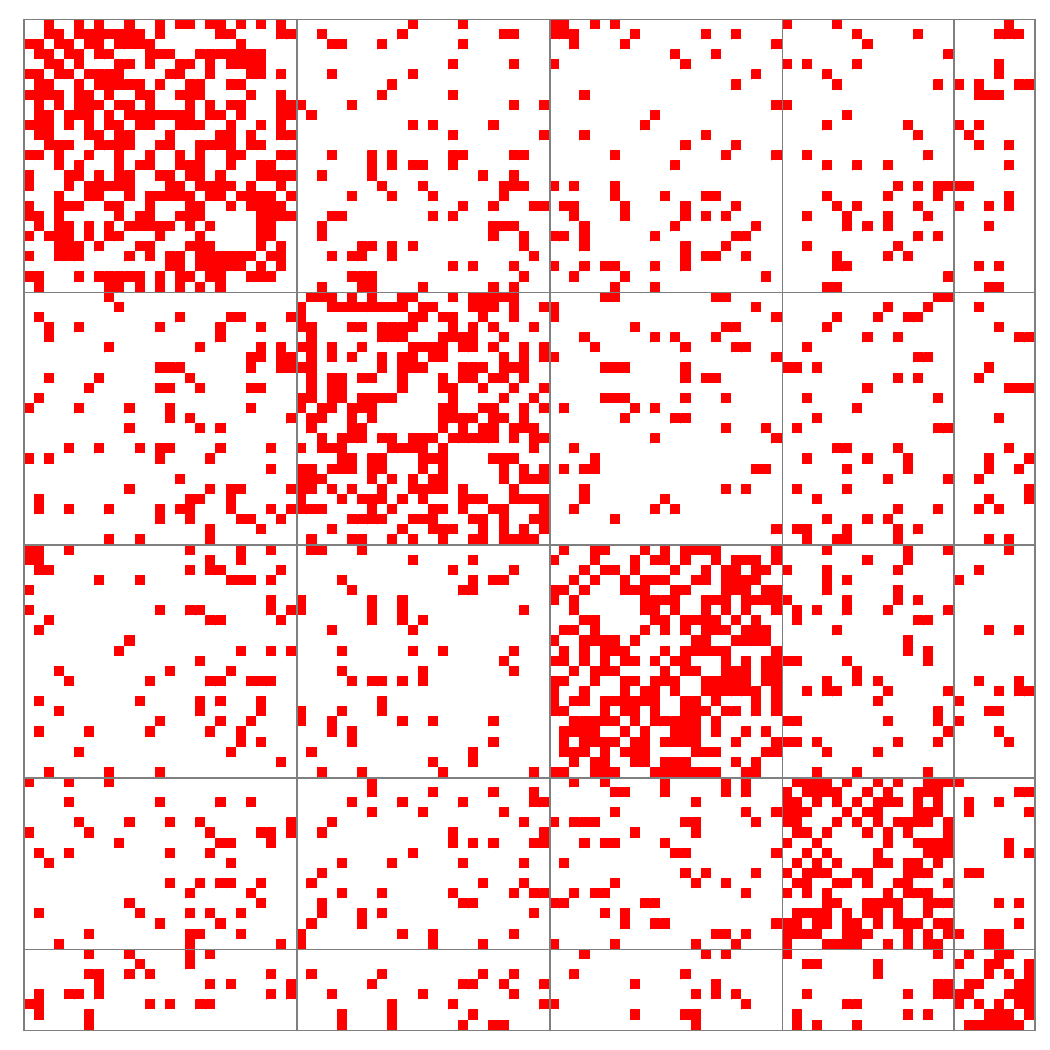}}
    \subfigure[Scenario 1: Graph.]        {\includegraphics[scale=0.43]{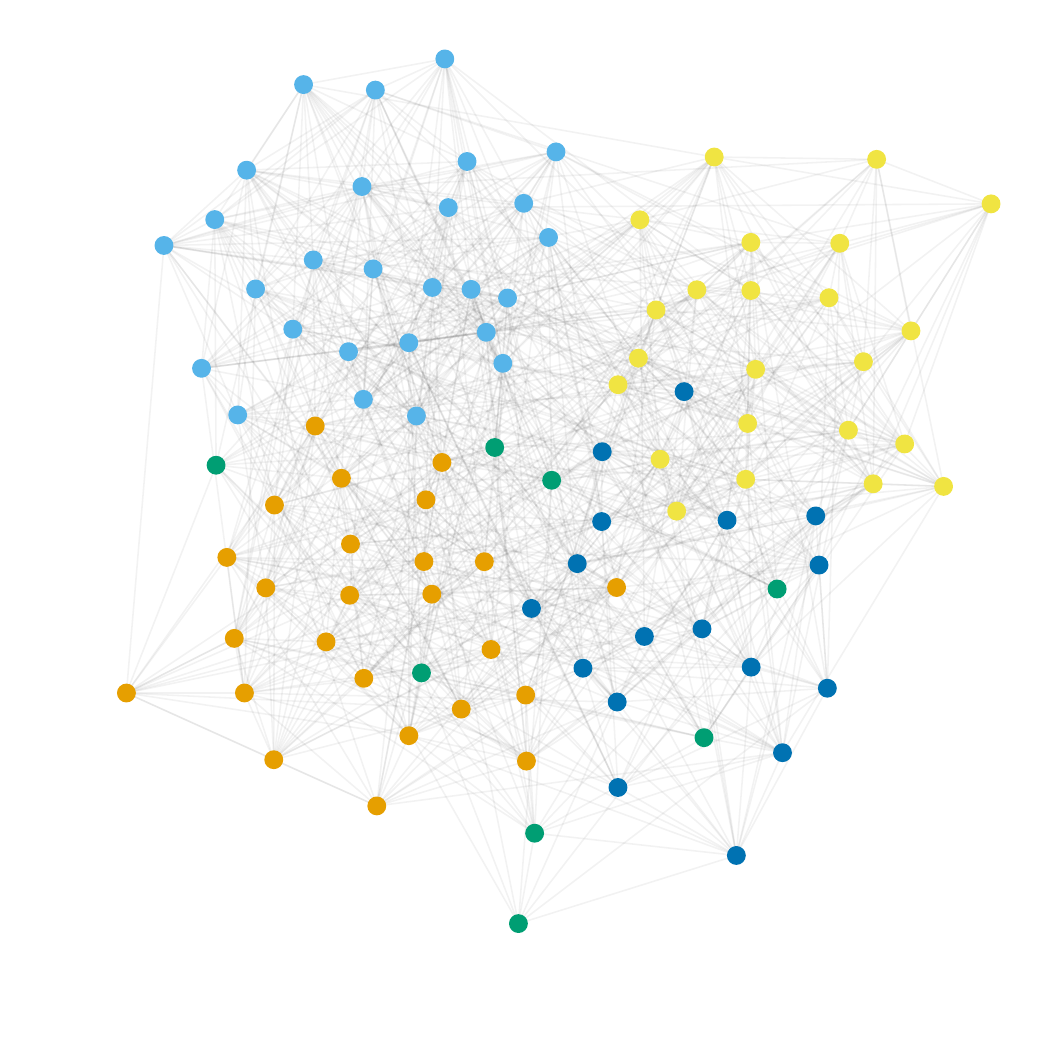}}
    \subfigure[Scenario 2: Socio matrix.] {\includegraphics[scale=0.43]{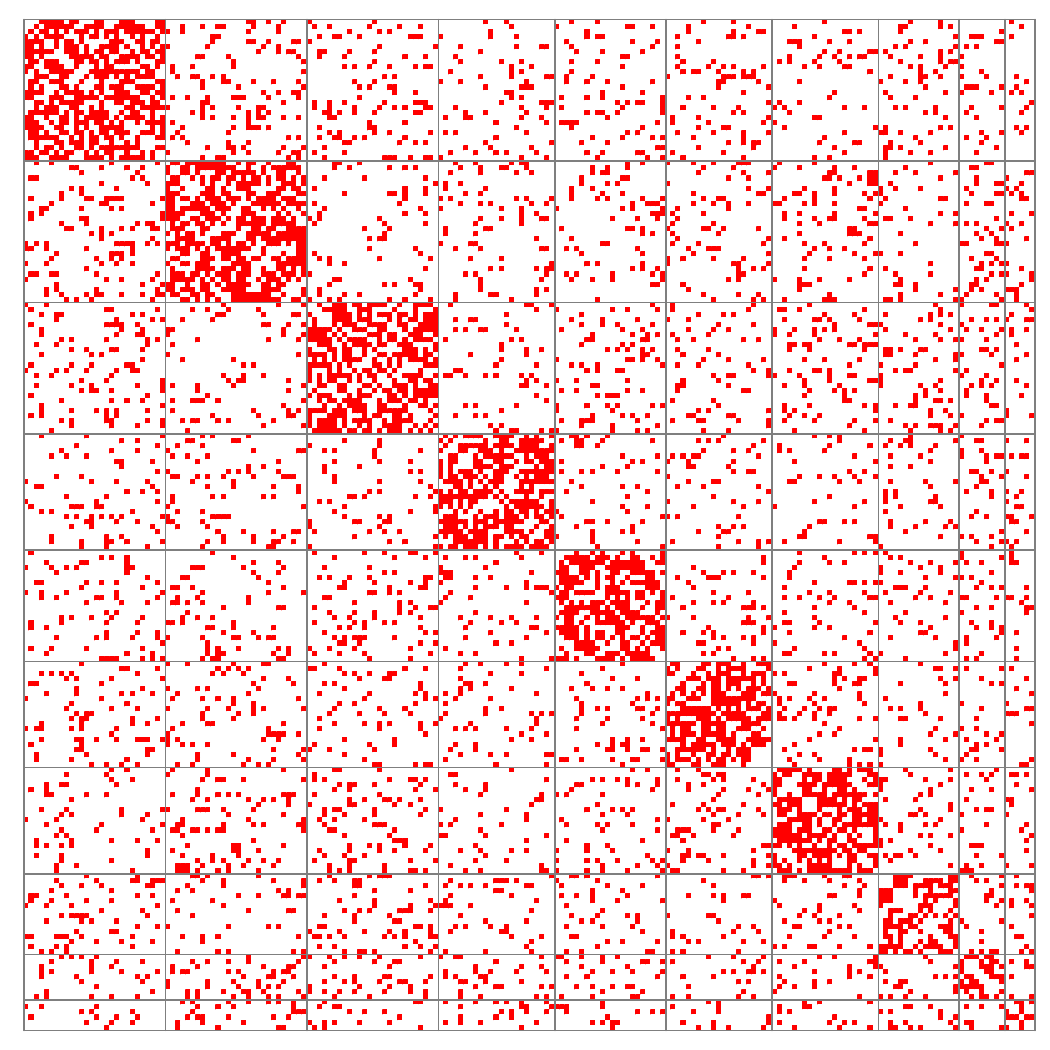}}
    \subfigure[Scenario 2: Graph.]        {\includegraphics[scale=0.43]{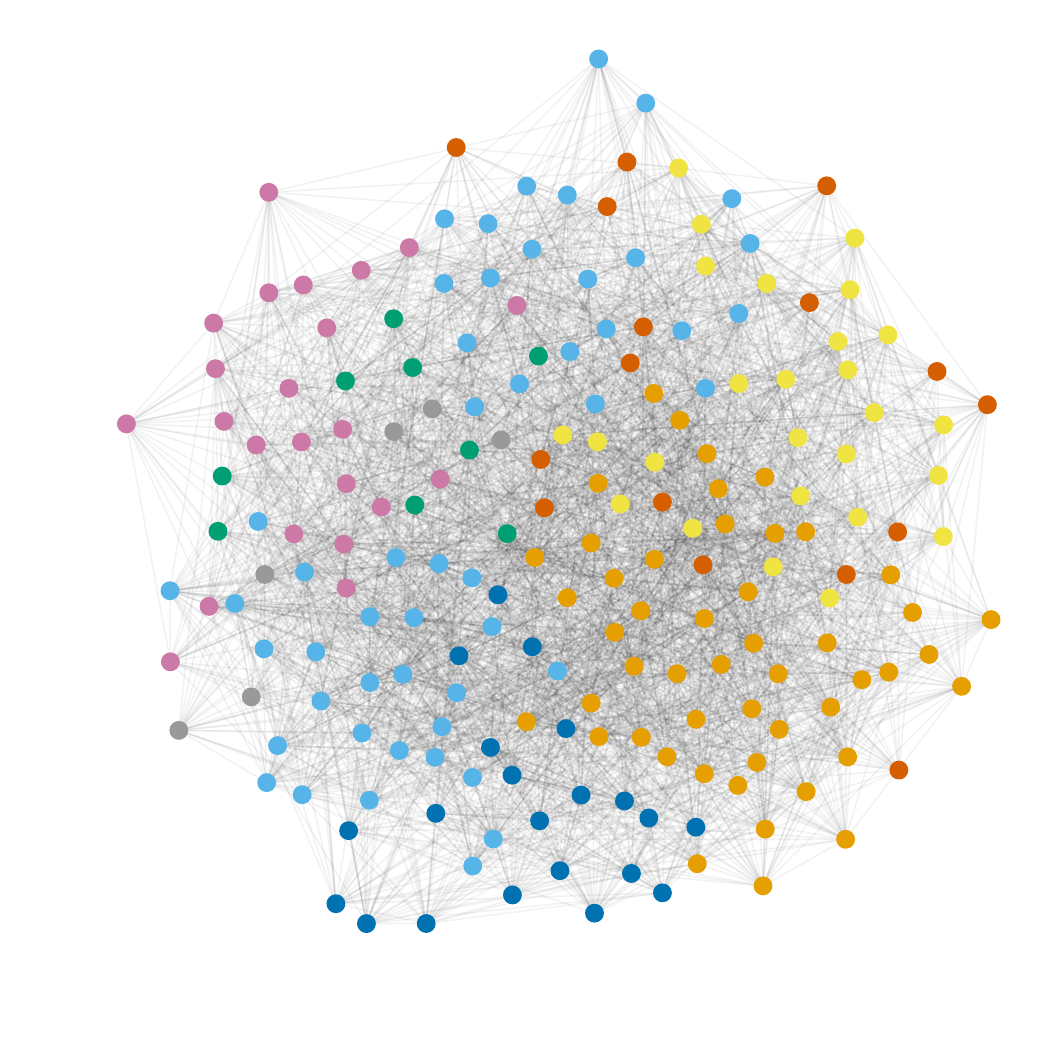}}
    \caption{Simulated data used to assess the performance of standard and hybrid class models.}
    \label{fig:simulation_data}
\end{figure}

In this simulation study, we generate two networks, each sampled from a Bernoulli distribution. The within-block probabilities are randomly drawn from 0.50, 0.52, 0.54, 0.56, 0.58, 0.60, and the between-block probabilities from 0.10, 0.11, 0.12, 0.13, 0.14, 0.15. These probabilities are selected to ensure that the separation between communities is subtle, making community detection nontrivial due to the relatively small gap between within- and between-block connection probabilities. The two networks have sizes \(I = 100\) with \(K = 5\) communities and \(I = 200\) with \(K = 10\) communities, which we refer to as Scenario 1 and Scenario 2, respectively. These settings allow us to evaluate model performance under different sample sizes and numbers of communities, thus testing scalability and robustness. In both cases, community sizes are randomly generated to produce blocks of varying sizes, enabling us to assess the models’ ability to recover heterogeneous and unbalanced cluster structures. Visual representations of both networks are shown in Figure~\ref{fig:simulation_data}, including graph layouts and sociomatrices. In the network graphs, ground-truth communities are indicated by vertex colors, while in the sociomatrices, they are delineated using horizontal and vertical lines.

We now apply the evaluation metrics described in Section \ref{sec:evaluation} to assess the performance of the different class models across clustering priors. Table \ref{tab:model_comparison_scenario_1} presents the results for Scenario 1, where the true number of clusters is 5. In the CM model, the DP prior stands out with the lowest posterior mean VI distance to the true partition ($\text{VI}_{\mathbf{z}_0} = 0.166$), the highest ARI (0.955), and a median number of clusters matching the ground truth. The PY prior also performs well, achieving low VI, FDR, and FNR values, though the DP prior exhibits slightly better fit and predictive accuracy. The GN prior, while producing adequate clustering accuracy, exhibits elevated uncertainty as reflected by the highest $\text{VI}_{\mathbf{z}_b}$ (0.934). Within the CDM model, the DP and PY priors offer the most accurate and stable clustering outcomes, with the lowest $\text{VI}_{\mathbf{z}_0}$ (0.126 and 0.130), highest ARI (0.968 and 0.967), and minimal FDR (0.005), while also recovering the correct number of clusters. These priors further yield the lowest WAIC. The CBM model, although competitive, offers no substantial improvements over CDM in clustering accuracy or prediction. Overall, results from Scenario 1 indicate that the CDM model under the DP or PY prior most effectively recovers the true cluster structure while maintaining low uncertainty and strong predictive ability.

\begin{table}[!ht]
\centering
\caption{Model comparison based on clustering and predictive metrics.}
{\small
\begin{tabular}{llcccccccc}
\toprule
Model & Prior & $\text{VI}_{\mathbf{z}_0}$ & $\text{VI}_{\mathbf{z}_b}$ & $H$ & ARI & FDR & FNR & WAIC \\
\midrule
CM  & DM & 0.326 & 0.394 & 4 & 0.905 & 0.122 & 0.013 & 4715 \\
CM  & DP & 0.166 & 0.482 & 5 & 0.955 & 0.054 & 0.011 & 4850 \\
CM  & PY & 0.220 & 0.447 & 4 & 0.934 & 0.079 & 0.012 & 4699 \\
CM  & GN & 0.190 & 0.934 & 5 & 0.933 & 0.071 & 0.011 & 4888 \\
\midrule
CDM & DM & 0.388 & 0.411 & 4 & 0.901 & 0.126 & 0.015 & 4766 \\
CDM & DP & 0.126 & 0.377 & 5 & 0.968 & 0.005 & 0.042 & 4861 \\
CDM & PY & 0.130 & 0.362 & 5 & 0.967 & 0.005 & 0.042 & 4859 \\
CDM & GN & 0.145 & 0.431 & 5 & 0.960 & 0.005 & 0.052 & 4865 \\
\midrule
CBM & DM & 0.314 & 0.327 & 4 & 0.907 & 0.120 & 0.012 & 4721 \\
CBM & DP & 0.154 & 0.425 & 6 & 0.961 & 0.007 & 0.050 & 4885 \\
CBM & PY & 0.169 & 0.442 & 6 & 0.958 & 0.007 & 0.054 & 4885 \\
CBM & GN & 0.199 & 0.515 & 6 & 0.944 & 0.007 & 0.073 & 4893 \\
\bottomrule
\end{tabular}
}
\caption{Comparison of model performance across clustering priors for each class model type in Scenario 1. Performance is evaluated using the posterior mean of the Variation of Information (VI) distance from the true partition (denoted $\text{VI}_{\mathbf{z}_0}$), the VI distance between the estimated partition $\mathbf{z}$ and the edge $\mathbf{z}_b$ of the 95\% credible ball (denoted $\text{VI}_{\mathbf{z}_b}$), the posterior median number of non-empty clusters $H$, and the posterior mean of the Adjusted Rand Index (ARI), the False Discovery Rate (FDR), and the False Negative Rate (FNR). Additionally, model fit and predictive accuracy are assessed using the Widely Applicable Information Criterion (WAIC).}
\label{tab:model_comparison_scenario_1}
\end{table}

\begin{table}[!ht]
\centering
\caption{Model comparison based on clustering and predictive metrics.}
{\small
\begin{tabular}{llcccccccc}
\toprule
Model & Prior & $\text{VI}_{\mathbf{z}_0}$ & $\text{VI}_{\mathbf{z}_b}$ & $H$ & ARI & FDR & FNR & WAIC \\
\midrule
CM  & DM & 1.226 & 0.190 & 5 & 0.516 & 0.569 & 0.012 & 18353 \\
CM  & DP & 1.569 & 1.647 & 4 & 0.343 & 0.683 & 0.015 & 17772 \\
CM  & PY & 0.572 & 0.625 & 7 & 0.805 & 0.270 & 0.017 & 18429 \\
CM  & GN & 1.069 & 0.952 & 5 & 0.612 & 0.481 & 0.020 & 18144 \\
\midrule
CDM & DM & 1.168 & 0.278 & 5 & 0.598 & 0.496 & 0.018 & 18075  \\
CDM & DP & 0.389 & 0.465 & 11 & 0.949 & 0.034 & 0.053 & 18008 \\
CDM & PY & 0.403 & 0.469 & 12 & 0.947 & 0.033 & 0.056 & 18045 \\
CDM & GN & 0.428 & 0.505 & 12 & 0.943 & 0.037 & 0.060 & 18098 \\
\midrule
CBM & DM & 1.212 & 0.210 & 5 & 0.519 & 0.565 & 0.014 & 18181 \\
CBM & DP & 0.959 & 0.611 & 8 & 0.753 & 0.298 & 0.092 & 18337 \\
CBM & PY & 0.969 & 0.652 & 8 & 0.752 & 0.297 & 0.096 & 18343 \\
CBM & GN & 0.993 & 0.707 & 8 & 0.686 & 0.376 & 0.092 & 18637 \\
\bottomrule
\end{tabular}
}
\caption{Comparison of model performance across clustering priors for each class model type in Scenario 2. Refer to the caption of Table \ref{tab:model_comparison_scenario_1} for detailed descriptions of the performance metrics reported.}
\label{tab:model_comparison_scenario2}
\end{table}

We now evaluate model performance for Scenario~2, in which the true number of clusters is 10. The results in Table~\ref{tab:model_comparison_scenario2} reveal substantial differences across model types and clustering priors. For the CM model, the PY prior offers the best performance, achieving the lowest posterior mean VI distance to the true partition ($\text{VI}_{\mathbf{z}_0} = 0.572$), the highest ARI (0.805), and a median number of clusters (7) that more closely approximates the true value compared to other priors. In contrast, the DP prior underestimates the number of clusters (median $H = 4$) and exhibits the worst clustering accuracy, with $\text{VI}_{\mathbf{z}_0} = 1.569$ and ARI = 0.343. The GN prior improves over DP and DM in both VI and ARI, but still underestimates $H$. The CDM model shows a marked improvement in recovering the true structure. All three nonparametric priors (DP, PY, and GN) yield clustering solutions with $H$ close to or above the true value, along with excellent ARI scores (above 0.94) and low FDR and FNR. The DP and PY priors yield nearly indistinguishable results, with slightly better WAIC for the DP prior, suggesting a mild advantage in overall model fit. In the CBM model, all priors underestimate the number of clusters, with $H = 8$ under nonparametric priors and $H = 5$ under the DM prior. The DP and PY priors perform similarly, achieving ARI values around 0.75 and moderate FDR and FNR. The GN prior under CBM performs the worst among the three, with the highest FNR (0.092) and lowest ARI (0.686). In summary, the CDM model with either the DP or PY prior performs best in Scenario~2. 

\section{Illustration}\label{sec:illustration}

As in the previous section, we evaluate the proposed models using three real-world networks from distinct social contexts. The first is a novel dataset on the Colombian armed conflict, involving \$I = 93\$ municipalities. A link indicates that, between 2015 and 2023, both municipalities experienced the presence of the same armed group, at least three violent events, and coca cultivation areas of 100 hectares or more. Identifying communities in this network is very important for uncovering territorial clusters where violence and illicit economies intersect, offering insights for policy, peacebuilding, and drug control. The network combines two data sources: SIEVACAC records of violent events with identified perpetrators, and satellite-based estimates of coca cultivation from the SIMCI project. Annual adjacency matrices were built using co-occurrence criteria and aggregated into a consolidated network using average cultivated area as the inclusion threshold. This serves as our main case study.

The other two networks are well-established examples from the literature and serve as complementary benchmarks for our proposal. The second network represents social and familial ties among $I = 99$ households in a rural village in southern Karnataka, India \citep{salter2017latent}, where an edge denotes kinship or shared participation in religious activities. The third network includes $I = 193$ individuals who recently engaged in injection drug use at a shooting gallery, a high-risk environment for communal drug consumption \citep{weeks2002social}. In this case, nodes represent individuals and edges capture reported social ties, including shared drug use. The resulting binary, undirected graph was obtained by aggregating egocentric survey responses into a complete sociometric network.

\begin{figure}[!htb]
    \centering
    \subfigure[Conflict data: Socio matrix.] {\includegraphics[scale=0.29]{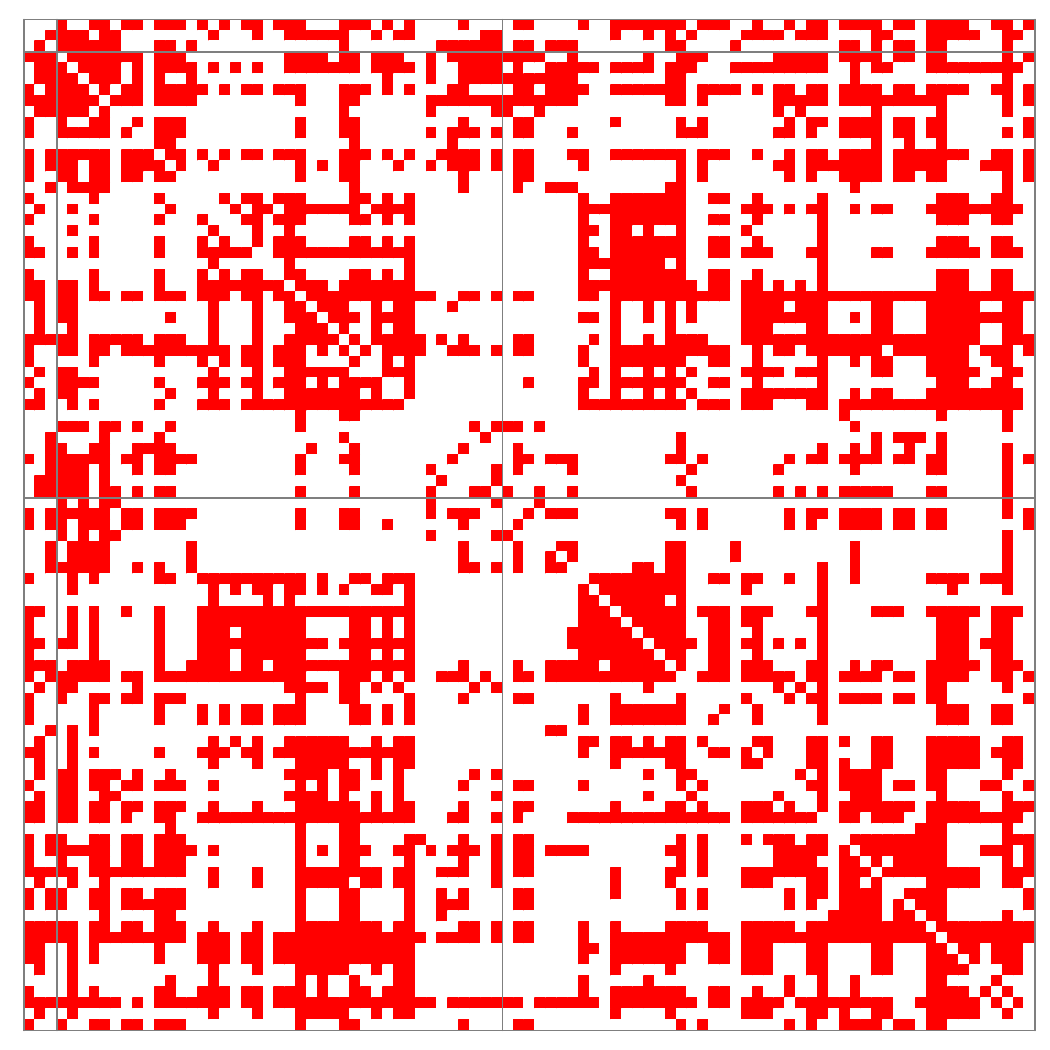}}
    \subfigure[Village data: Socio matrix.]  {\includegraphics[scale=0.29]{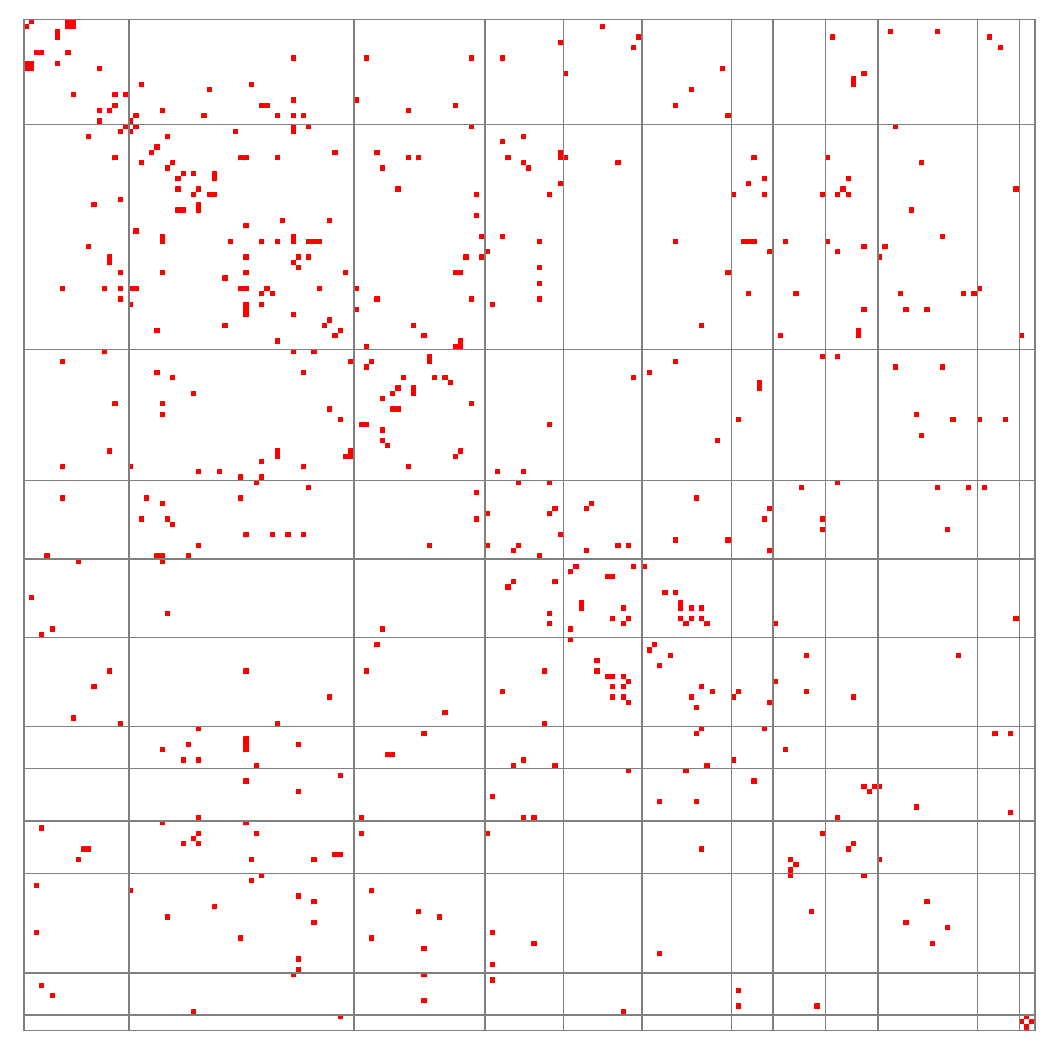}}
    \subfigure[Drug data: Socio matrix.]     {\includegraphics[scale=0.29]{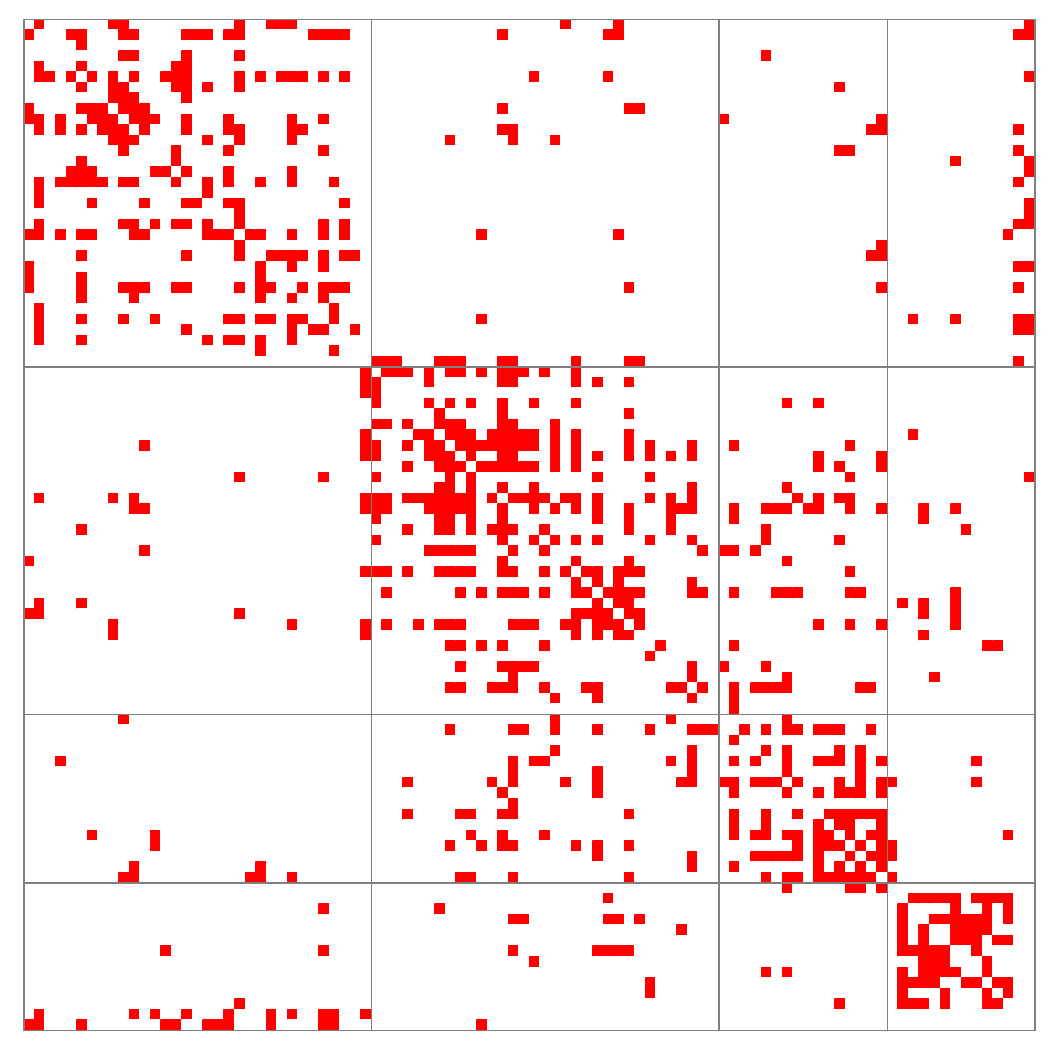}}
    \subfigure[conflict data: Graph.]        {\includegraphics[scale=0.29]{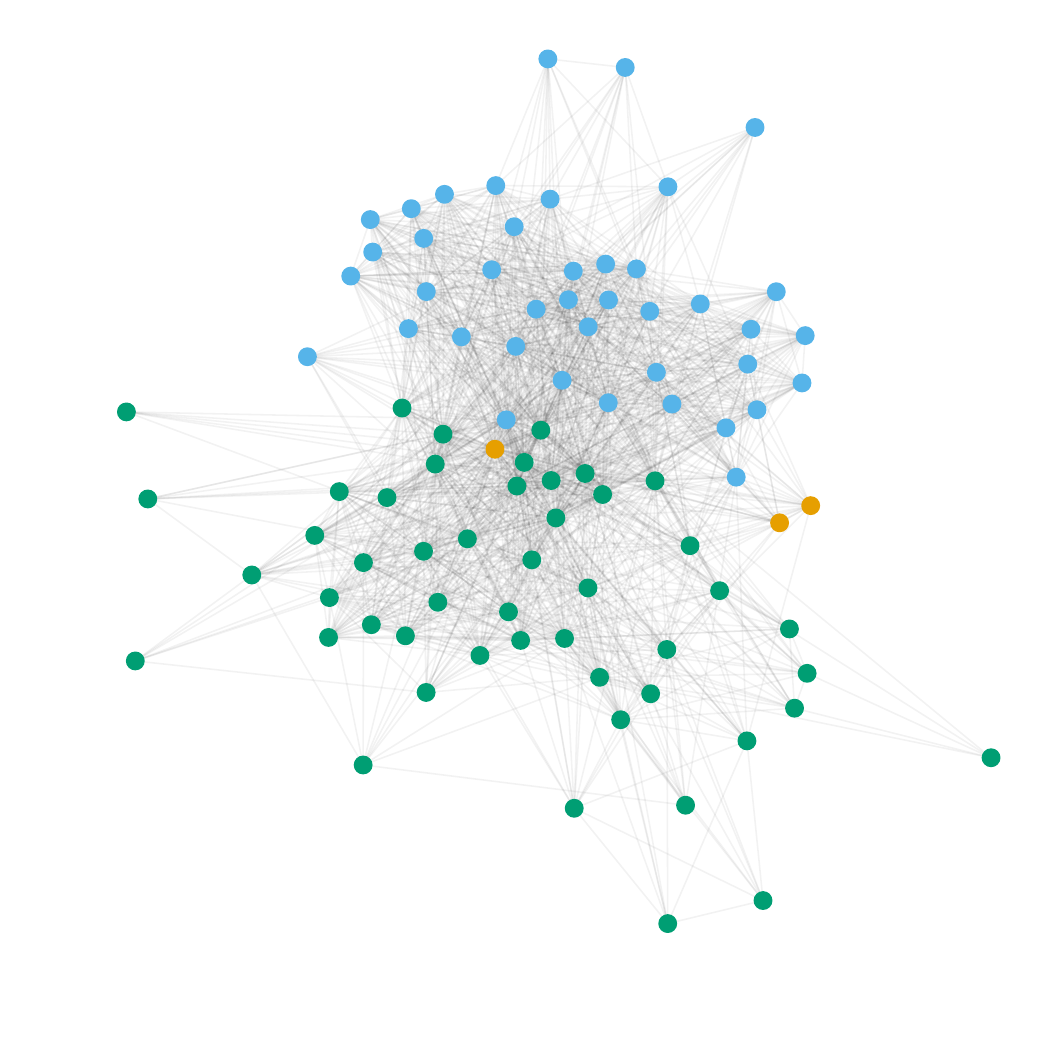}}
    \subfigure[Village data: Graph.]         {\includegraphics[scale=0.29]{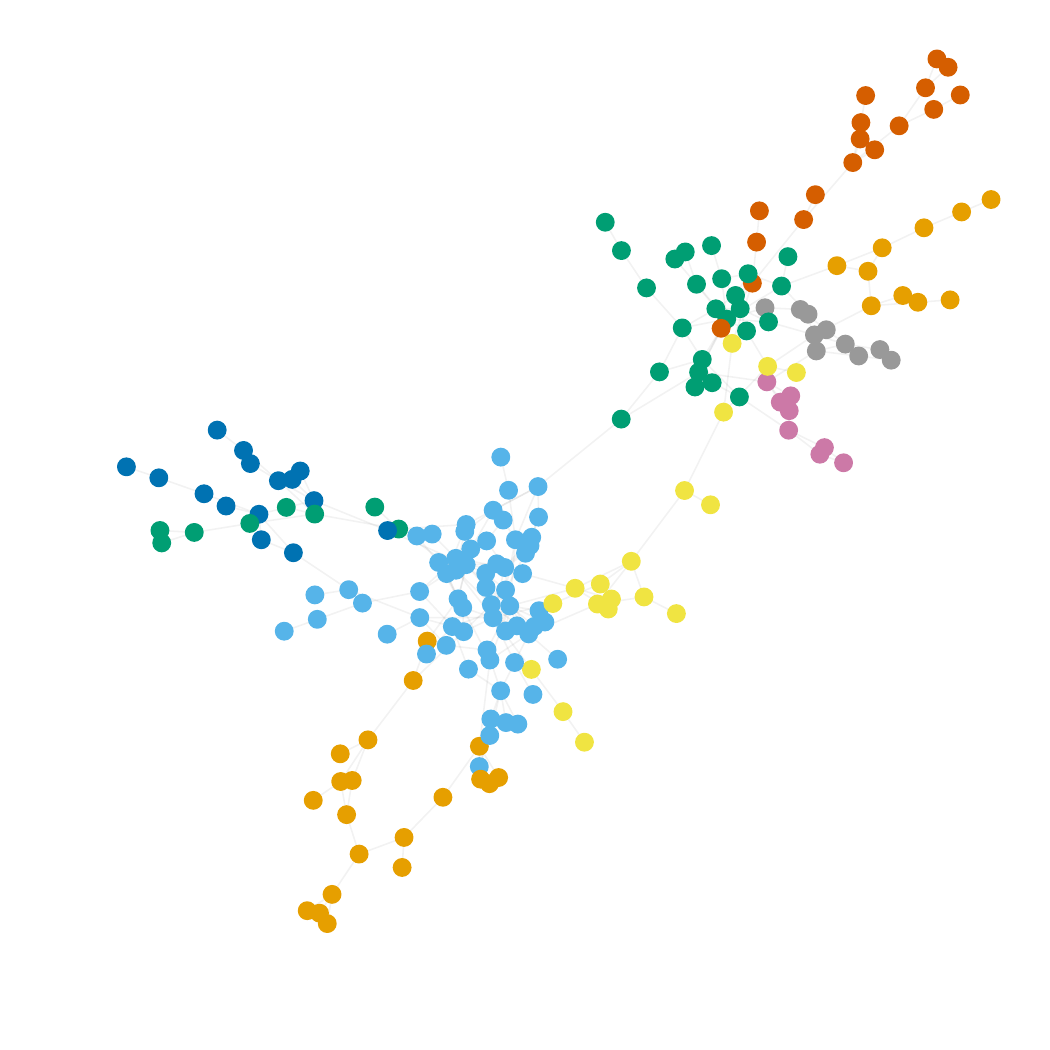}}
    \subfigure[Drug data: Graph.]            {\includegraphics[scale=0.29]{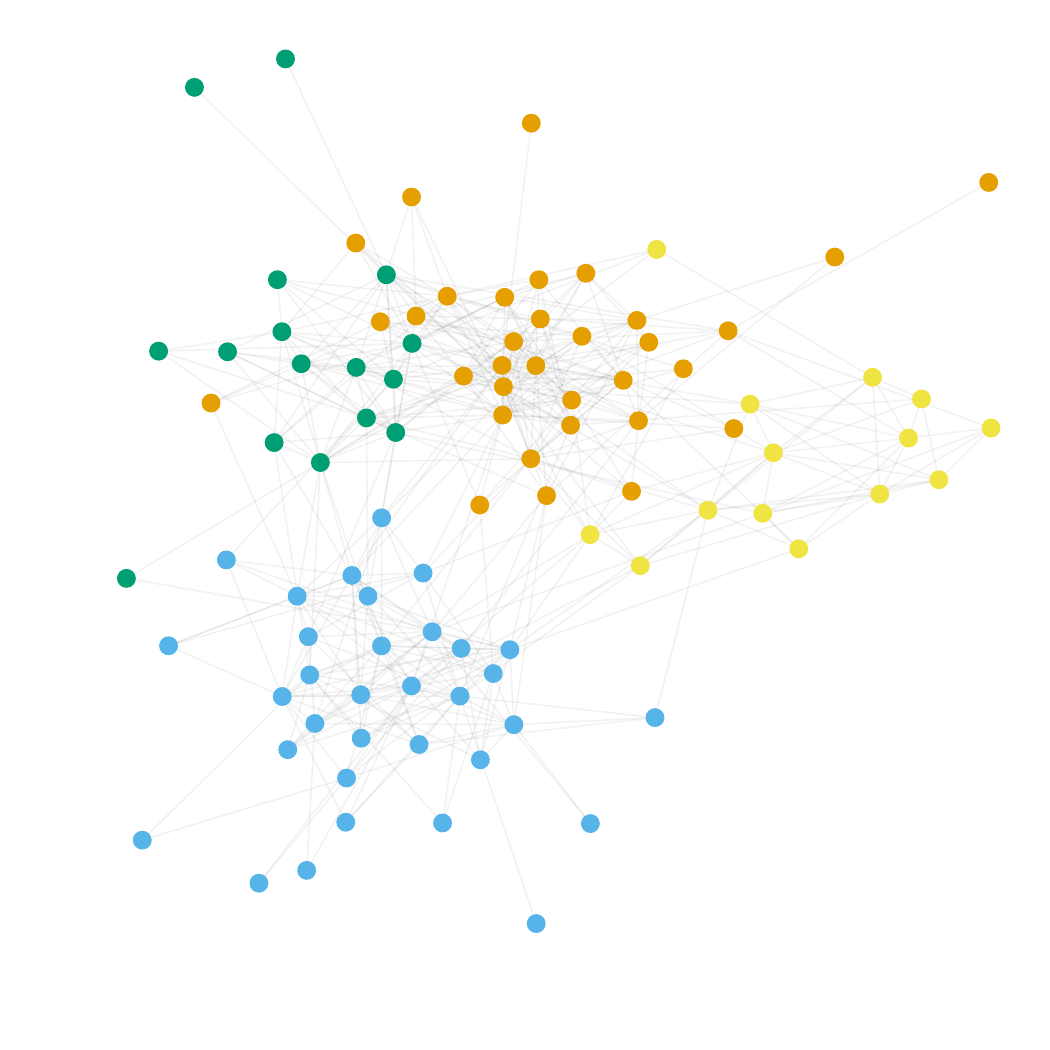}}
    \caption{Real data used to assess the performance of standard and hybrid class models.}
    \label{fig:illustration_data}
\end{figure}

Unlike the simulation study, these empirical applications pose a greater challenge, as the true underlying community structure is unknown. Figure~\ref{fig:illustration_data} presents the adjacency matrices and corresponding network visualizations for all datasets. In the network graphs, vertex colors reflect communities identified using the fast greedy modularity optimization algorithm \citep{clauset2004finding}, while in the sociomatrices, the same communities are indicated by horizontal and vertical separators.

We now apply the evaluation metrics described in Section~\ref{sec:evaluation} to assess the performance of the different class models across clustering priors. The results in Table~\ref{tab:realdata_pred_performance} illustrate the predictive performance of the three proposed model classes (CM, CDM, CBM) under four different clustering priors (DM, DP, PY, GN), evaluated on the Village data and the Drug data. The metrics considered are the Area Under the ROC Curve (AUC), Mean Squared Error (MSE), and Log-Loss. The primary objective here is to evaluate the predictive accuracy of the proposed models on real-world networks, where no ground-truth clustering is available. Consequently, the analysis focuses exclusively on predictive performance metrics, which directly assess each model’s ability to recover observed edges and allow for interpretable, consistent comparisons across model classes and priors. In contrast, clustering-specific metrics such as the median number of non-empty clusters, or model complexity criteria like the WAIC, are deliberately excluded, restricting the evaluation to predictive diagnostics that are robust and directly informative of model performance in out-of-sample prediction.

\begin{table}[!ht]
\centering
{\small
\begin{tabular}{llccccccccc}
\toprule
\multirow{2}{*}{Model} & \multirow{2}{*}{Prior} & \multicolumn{3}{c}{Conflict Data} & \multicolumn{3}{c}{Village Data} & \multicolumn{3}{c}{Drug Data} \\
\cmidrule(lr){3-5} \cmidrule(lr){6-8} \cmidrule(lr){9-11}
 & & AUC & MSE & Log-Loss & AUC & MSE & Log-Loss & AUC & MSE & Log-Loss \\
\midrule
CM   & DM & 0.731 & 0.186 & 0.558 & 0.893 & 0.062 & 0.209 & 0.897 & 0.014 & 0.058 \\
CM   & DP & 0.897 & 0.107 & 0.344 & 0.874 & 0.067 & 0.224 & 0.814 & 0.014 & 0.065 \\
CM   & PY & 0.900 & 0.101 & 0.320 & 0.855 & 0.069 & 0.233 & 0.786 & 0.014 & 0.067 \\
CM   & GN & 0.889 & 0.119 & 0.377 & 0.860 & 0.068 & 0.230 & 0.802 & 0.014 & 0.066 \\
CDM  & DM & 0.685 & 0.204 & 0.592 & 0.863 & 0.068 & 0.229 & 0.923 & 0.013 & 0.052 \\
CDM  & DP & 0.973 & 0.058 & 0.196 & 0.874 & 0.066 & 0.223 & 0.898 & 0.014 & 0.057 \\
CDM  & PY & 0.973 & 0.059 & 0.197 & 0.870 & 0.067 & 0.225 & 0.824 & 0.014 & 0.063 \\
CDM  & GN & 0.974 & 0.058 & 0.194 & 0.878 & 0.065 & 0.220 & 0.865 & 0.014 & 0.060 \\
CBM  & DM & 0.724 & 0.192 & 0.568 & 0.896 & 0.060 & 0.205 & 0.895 & 0.013 & 0.057 \\
CBM  & DP & 0.971 & 0.061 & 0.196 & 0.907 & 0.059 & 0.198 & 0.872 & 0.014 & 0.060 \\
CBM  & PY & 0.972 & 0.060 & 0.194 & 0.903 & 0.059 & 0.201 & 0.827 & 0.014 & 0.064 \\
CBM  & GN & 0.973 & 0.059 & 0.191 & 0.912 & 0.058 & 0.194 & 0.859 & 0.014 & 0.061 \\
\bottomrule
\end{tabular}
}
\caption{Comparison of model performance across clustering priors for each class model type applied to the Conflict, Village, and Drug networks. Metrics include the Area Under the ROC Curve (AUC), Mean Squared Error (MSE), and Log-Loss.}
\label{tab:realdata_pred_performance}
\end{table}

In the \textit{Conflict} network, which serves as our primary case study, the hybrid models \textsf{CDM} and \textsf{CBM} yield clear improvements over the baseline \textsf{CM} across all priors, with \textsf{CDM} (GN) and \textsf{CBM} (GN) achieving the best overall predictive performance. These results highlight the benefits of incorporating latent group-level structure to model the joint dynamics of armed violence and illicit economies. In contrast, \textsf{CM} underperforms significantly, particularly with the \textsf{DM} prior, reflecting its limited capacity to capture structural complexity in real-world settings. Similar trends are observed in the \textit{Village} and \textit{Drug} networks: \textsf{CBM} excels in the former, while \textsf{CDM} leads in the latter. Overall, \textsf{CDM} and \textsf{CBM} consistently outperform \textsf{CM} in predictive accuracy and calibration, confirming the strength of hybrid latent structure in capturing nuanced connectivity patterns across diverse relational settings.

Regarding the conflict data, the clustering reveals a detailed segmentation of the Colombian territory shaped by overlapping patterns of armed violence and illicit economies. Cluster 1 links municipalities in northern Antioquia, southern Bolívar, and the Pacific coast (e.g., Anorí, San Pablo, Guapi), underscoring guerrilla presence and established trafficking routes. Cluster 2 brings together coca-producing zones in Cauca and Putumayo (e.g., Briceño, Jamundí, Cartagena del Chairá), primarily influenced by ELN and FARC dissident factions. Cluster 3 spans areas such as Segovia, Morales (Bolívar), and several municipalities in Chocó and Córdoba, representing territories marked by illegal mining and post-demobilization violence. Cluster 6 includes eastern and border municipalities (e.g., Arauquita, Simití, Tibú) where ELN governance structures are particularly strong. These clusters not only capture regional proximity but also reflect cross-regional coordination and territorial control strategies among armed actors. As such, the resulting classification provides a valuable basis for informed peacebuilding, differentiated drug policy, and geographically targeted violence prevention initiatives.

\begin{figure}[!htb]
    \centering
    \subfigure[Density.]       {\includegraphics[scale=0.2]{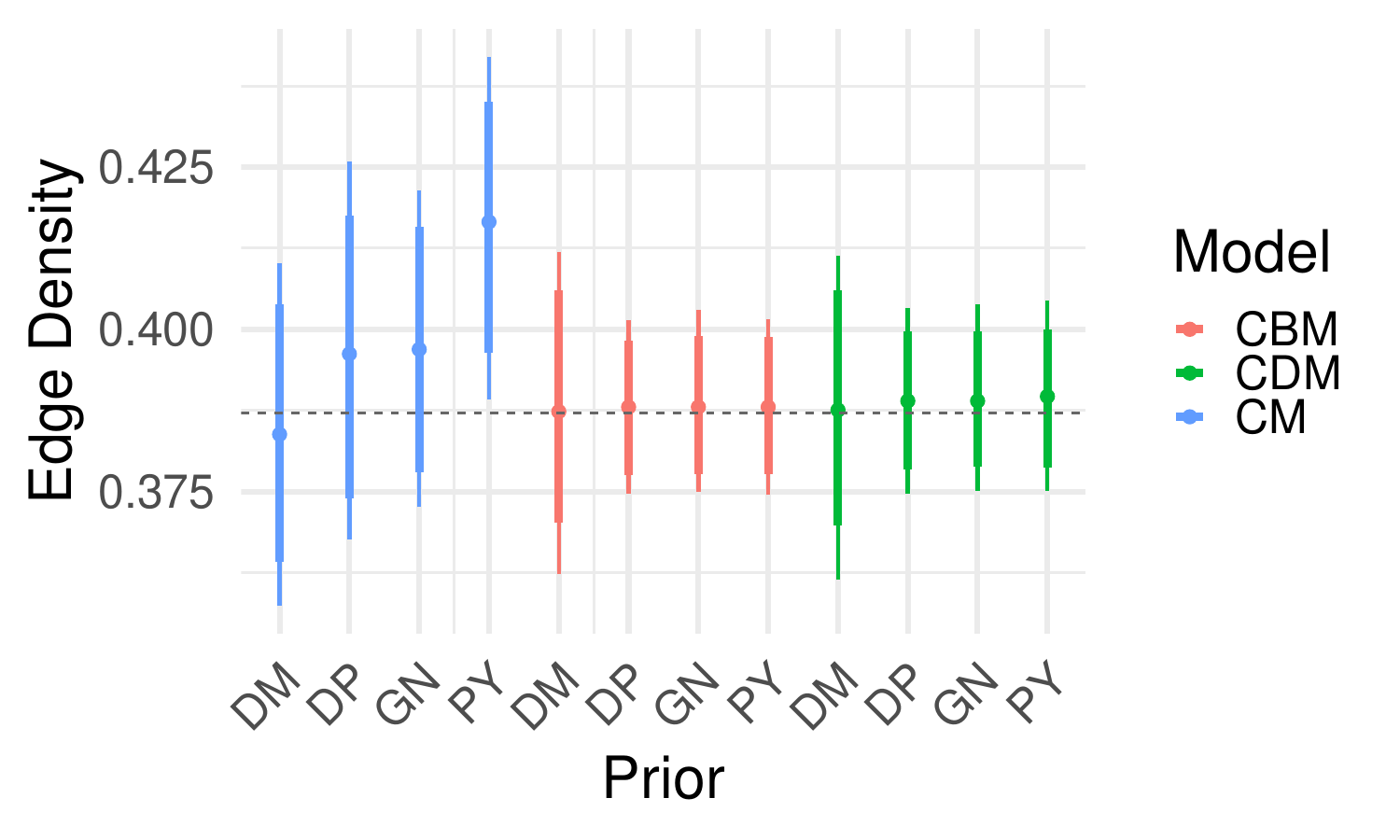}}
    \subfigure[Transitivity.]  {\includegraphics[scale=0.2]{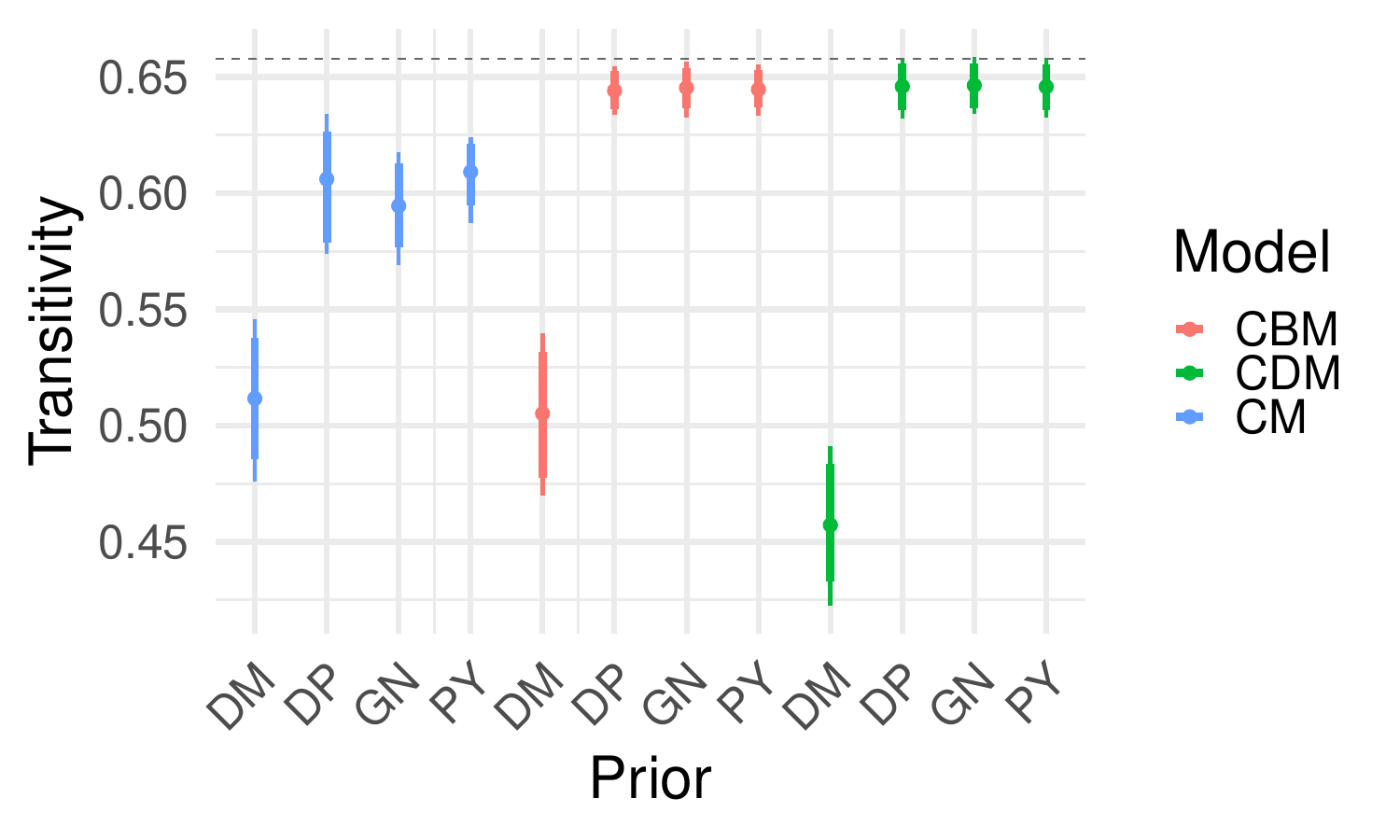}}
    \subfigure[Assortativity.] {\includegraphics[scale=0.2]{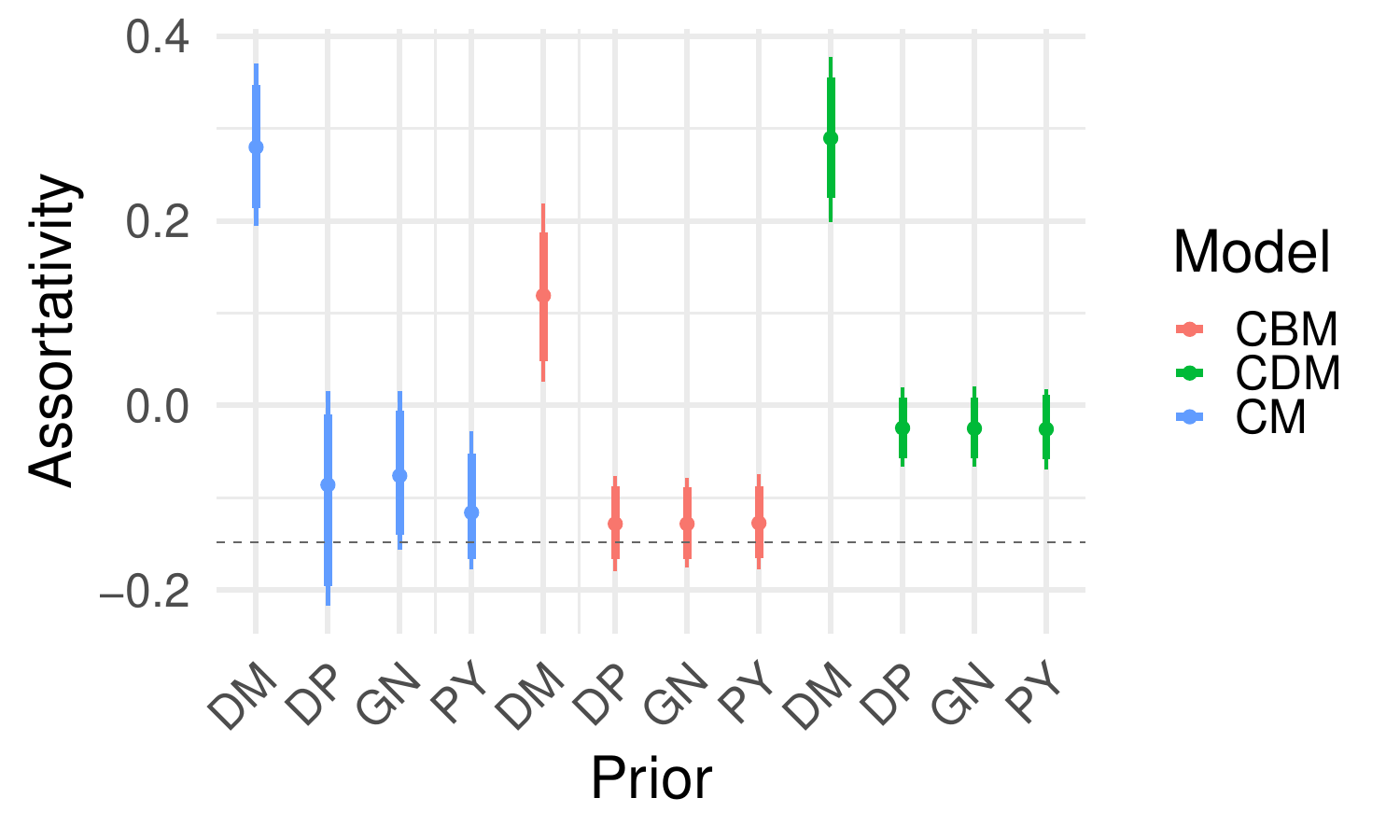}}
    \subfigure[Mean Degree.]   {\includegraphics[scale=0.2]{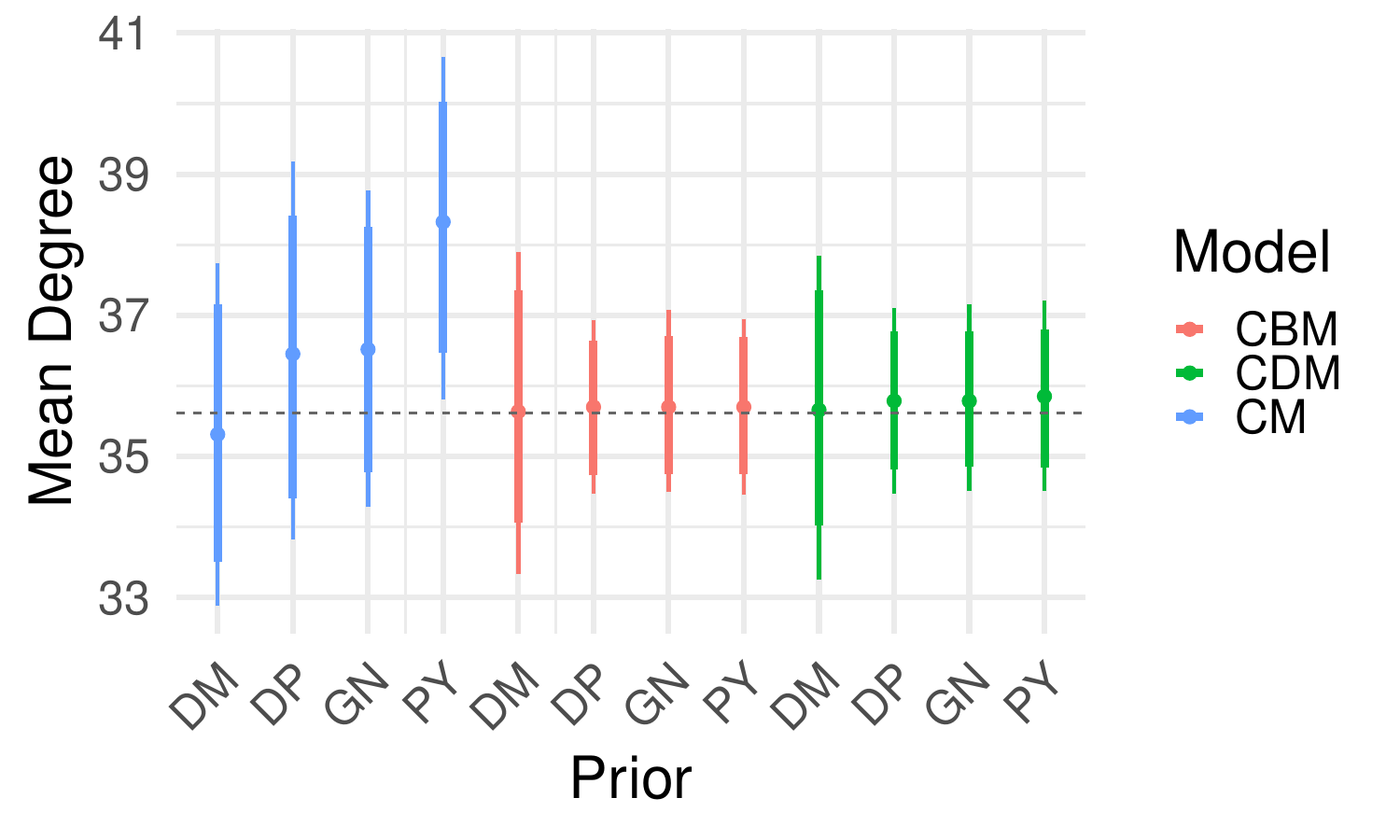}}
    \subfigure[SD Degree.]     {\includegraphics[scale=0.2]{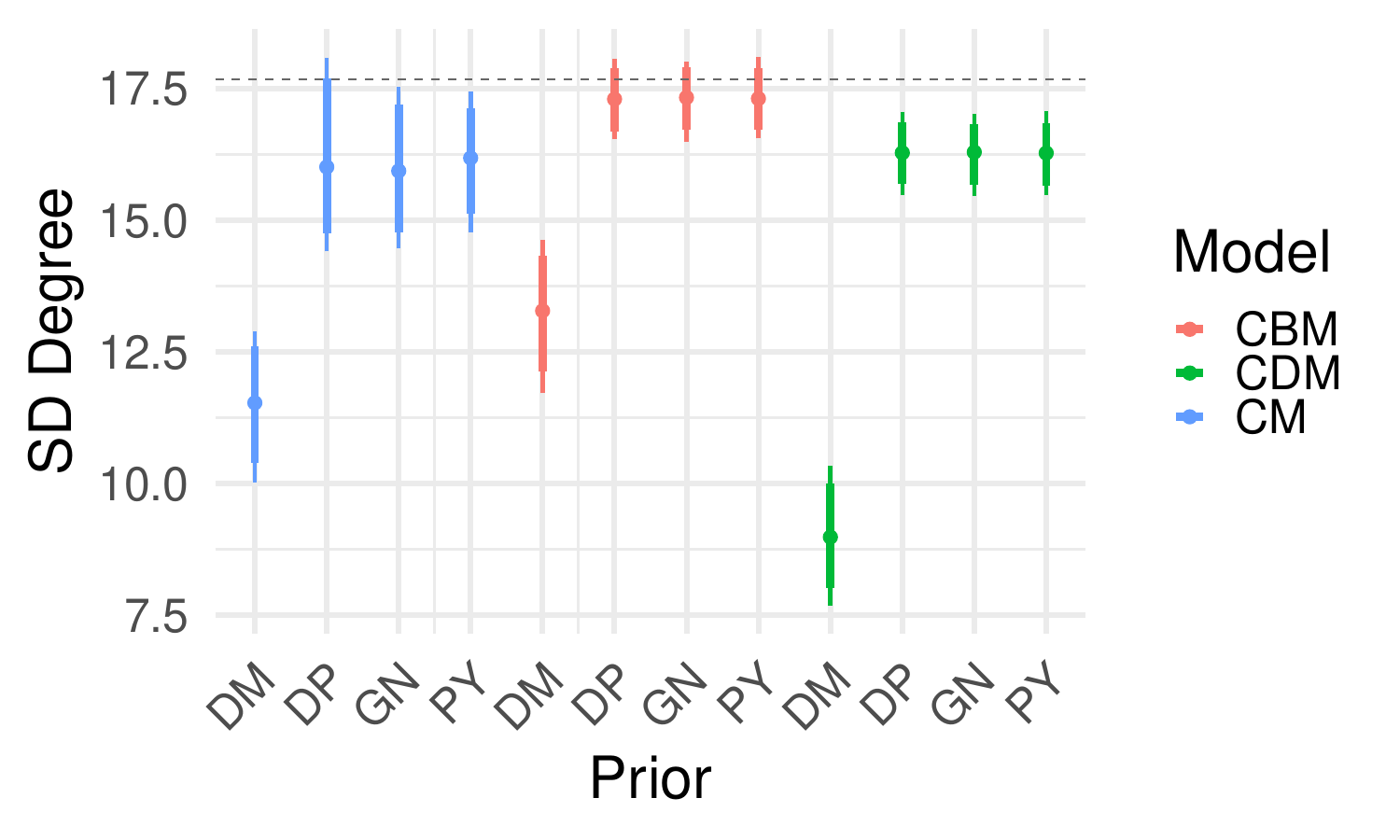}}
    \subfigure[Mean Distance.] {\includegraphics[scale=0.2]{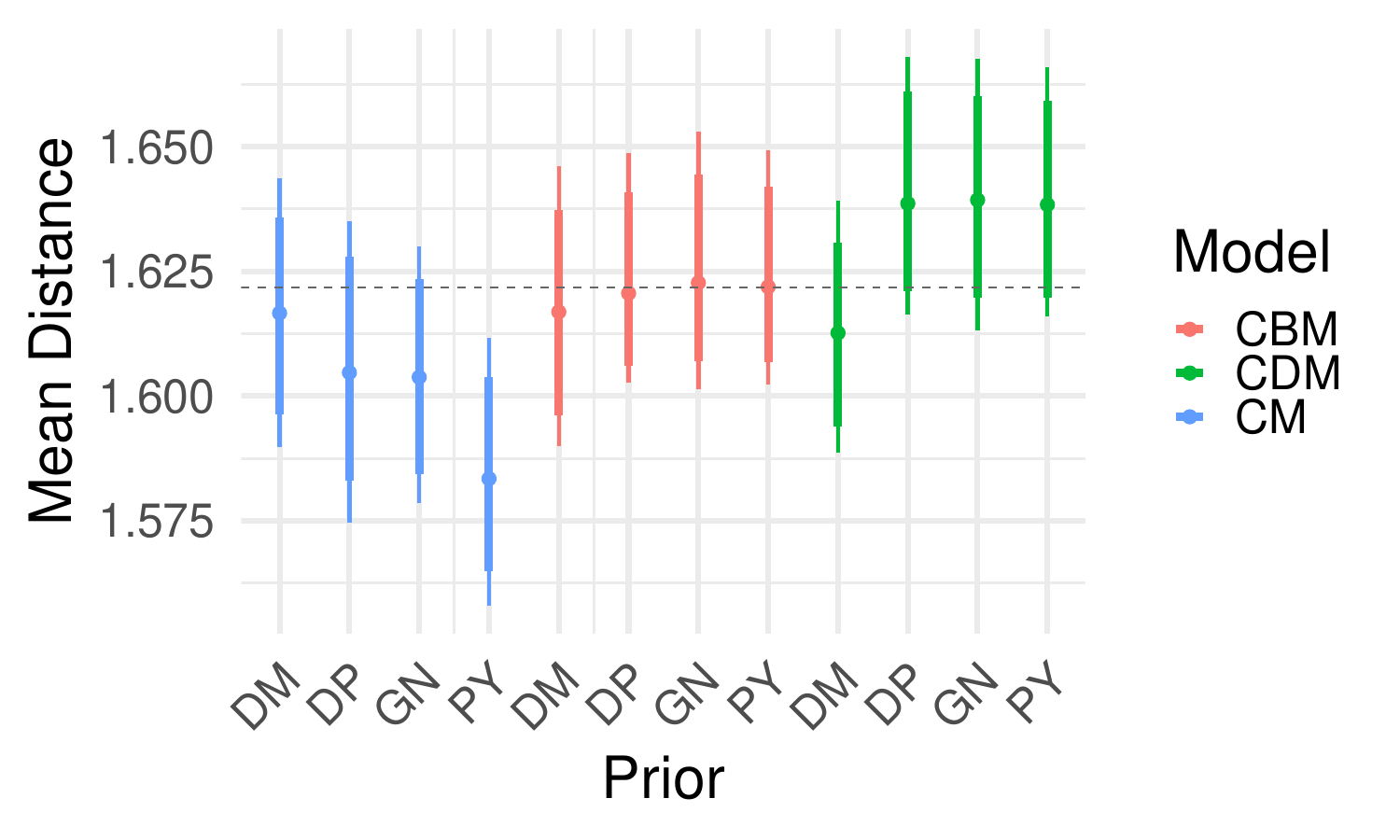}}
    \caption{Posterior predictive checks for key network statistics in the Conflict dataset.}
    \label{fig:test_statistics_conflict}
\end{figure}

\begin{figure}[!htb]
    \centering
    \subfigure[Density.]       {\includegraphics[scale=0.2]{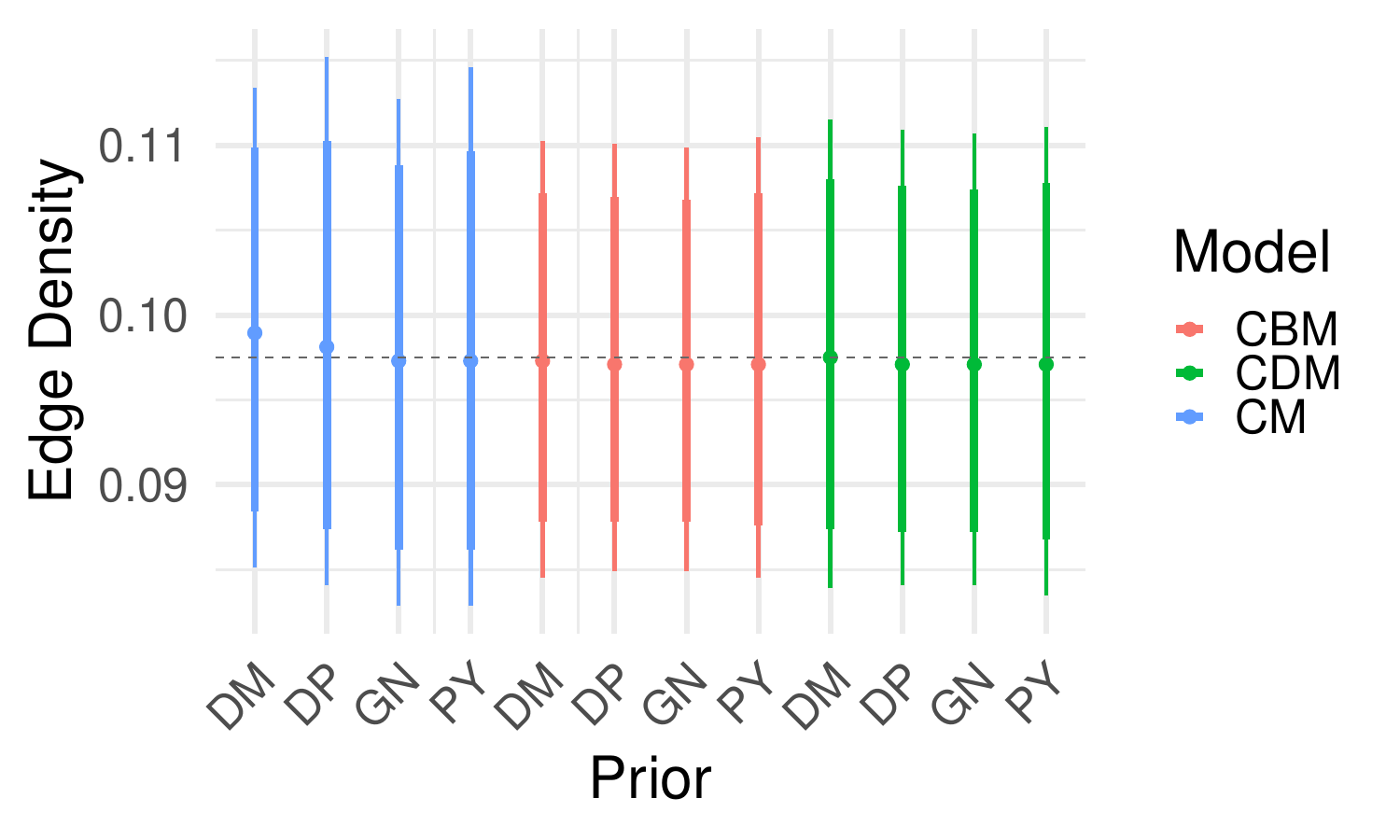}}
    \subfigure[Transitivity.]  {\includegraphics[scale=0.2]{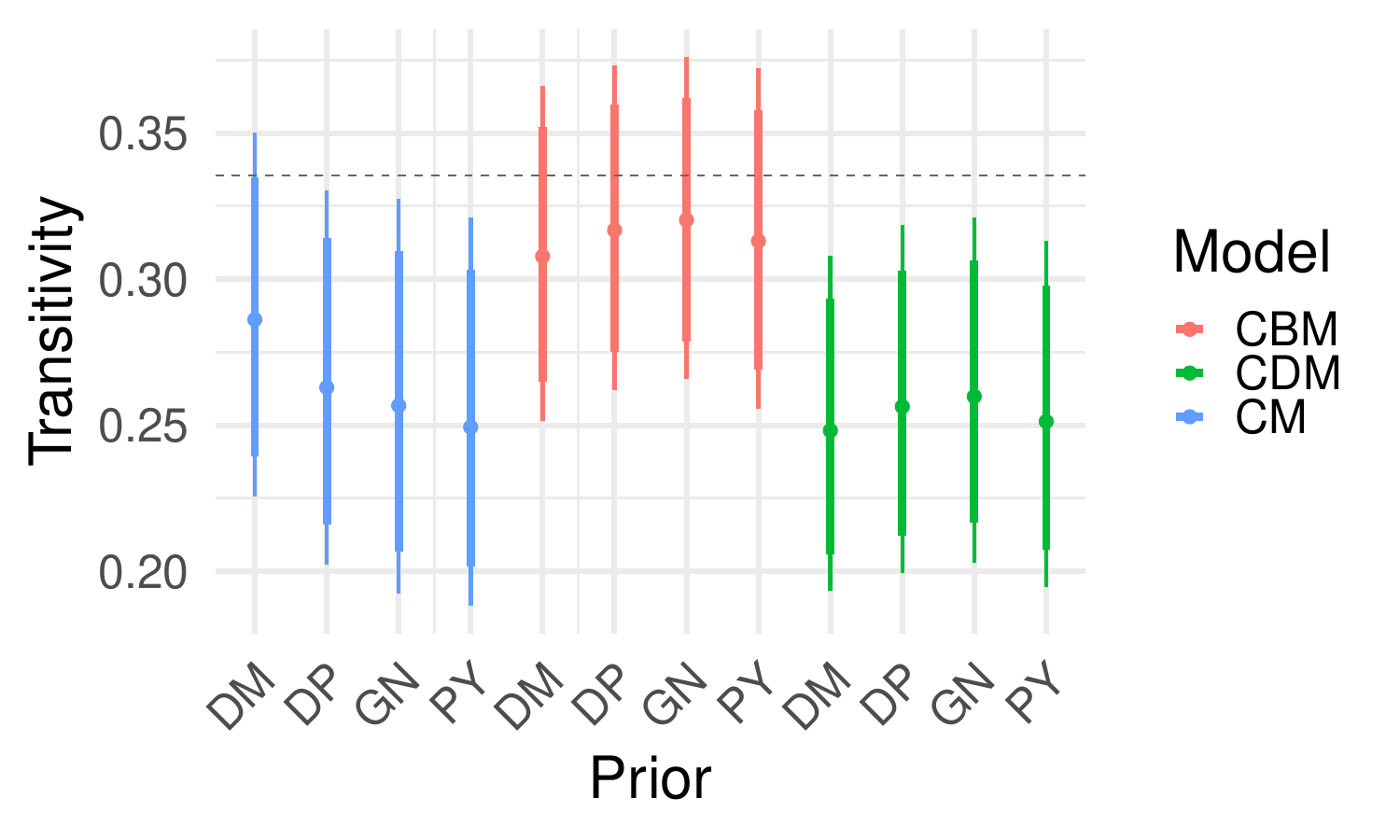}}
    \subfigure[Assortativity.] {\includegraphics[scale=0.2]{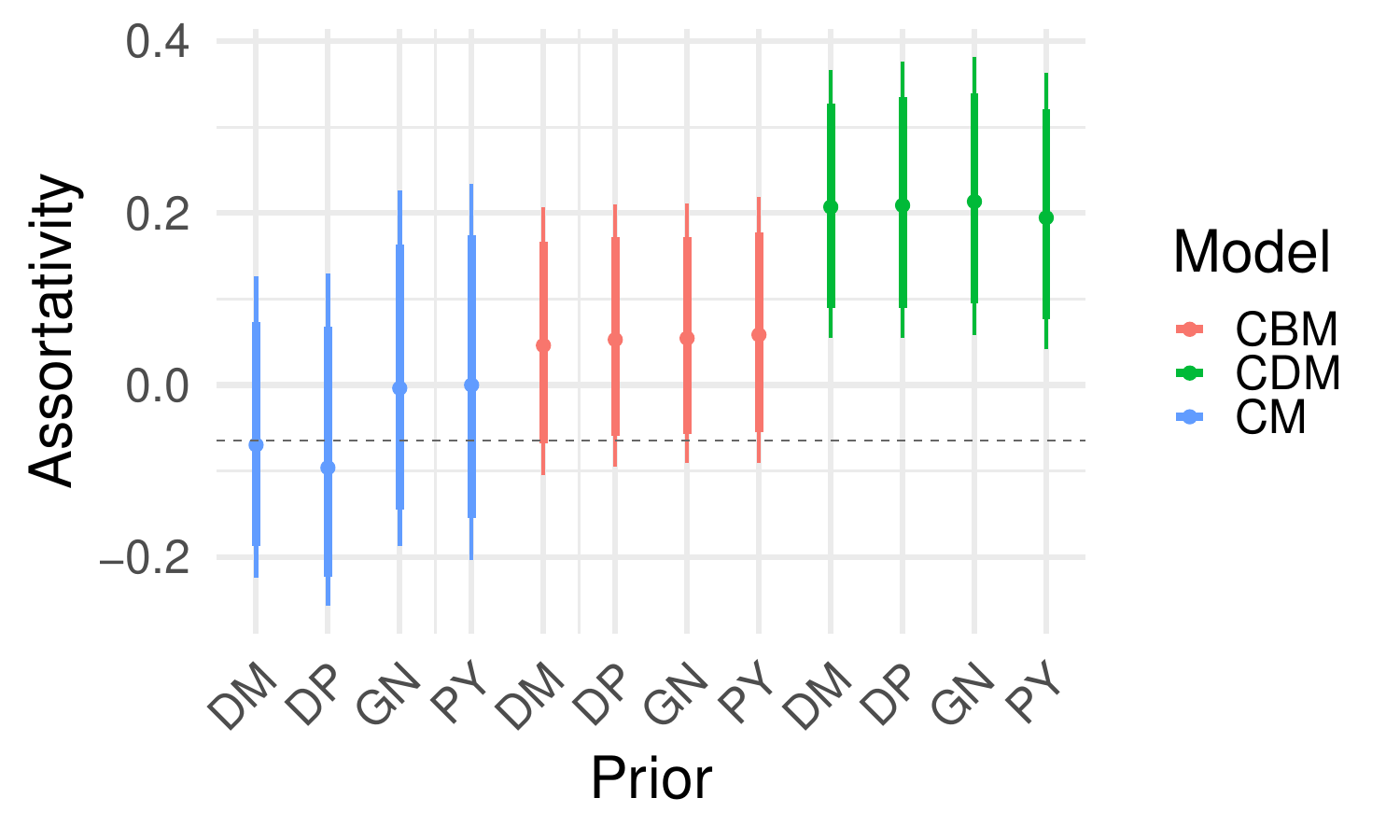}}
    \subfigure[Mean Degree.]   {\includegraphics[scale=0.2]{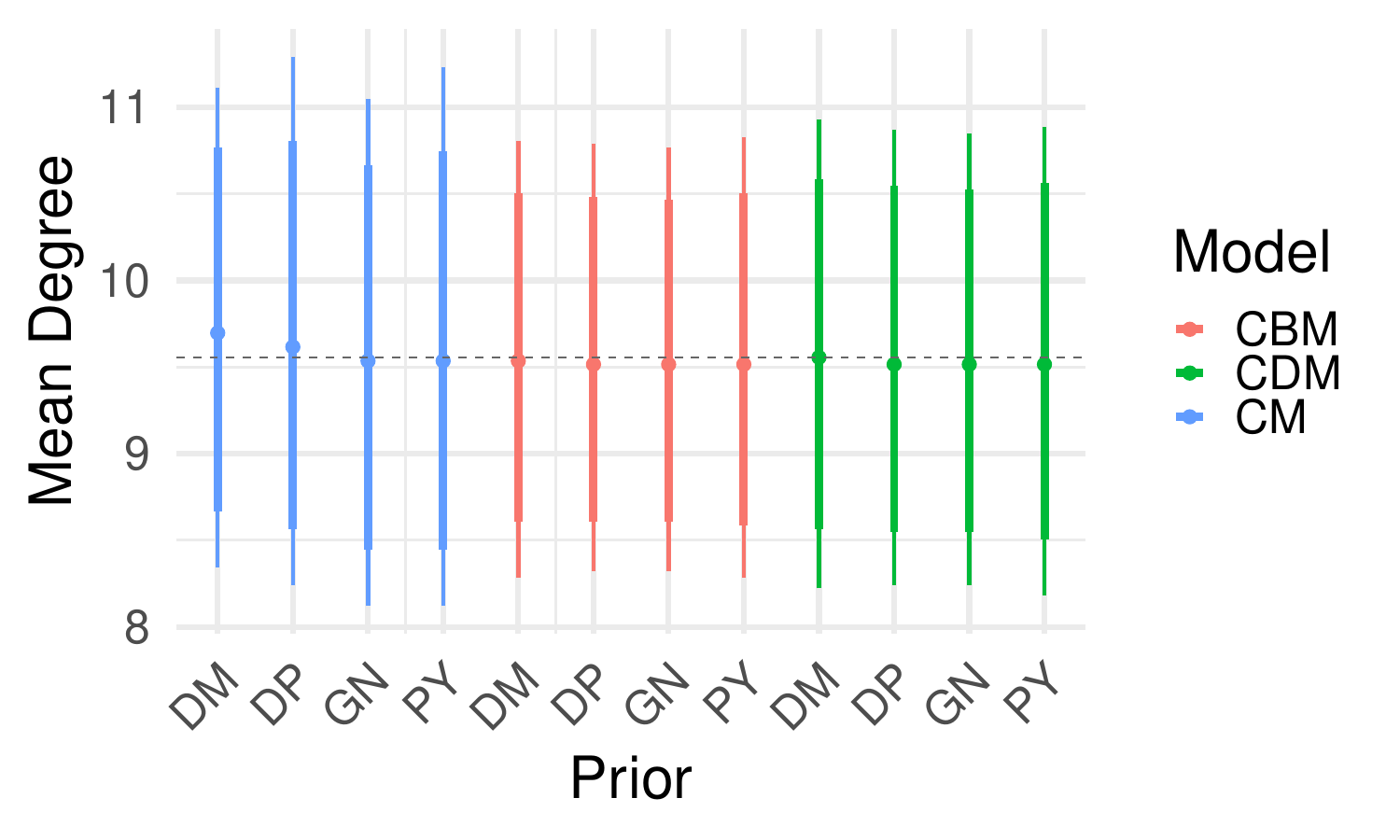}}
    \subfigure[SD Degree.]     {\includegraphics[scale=0.2]{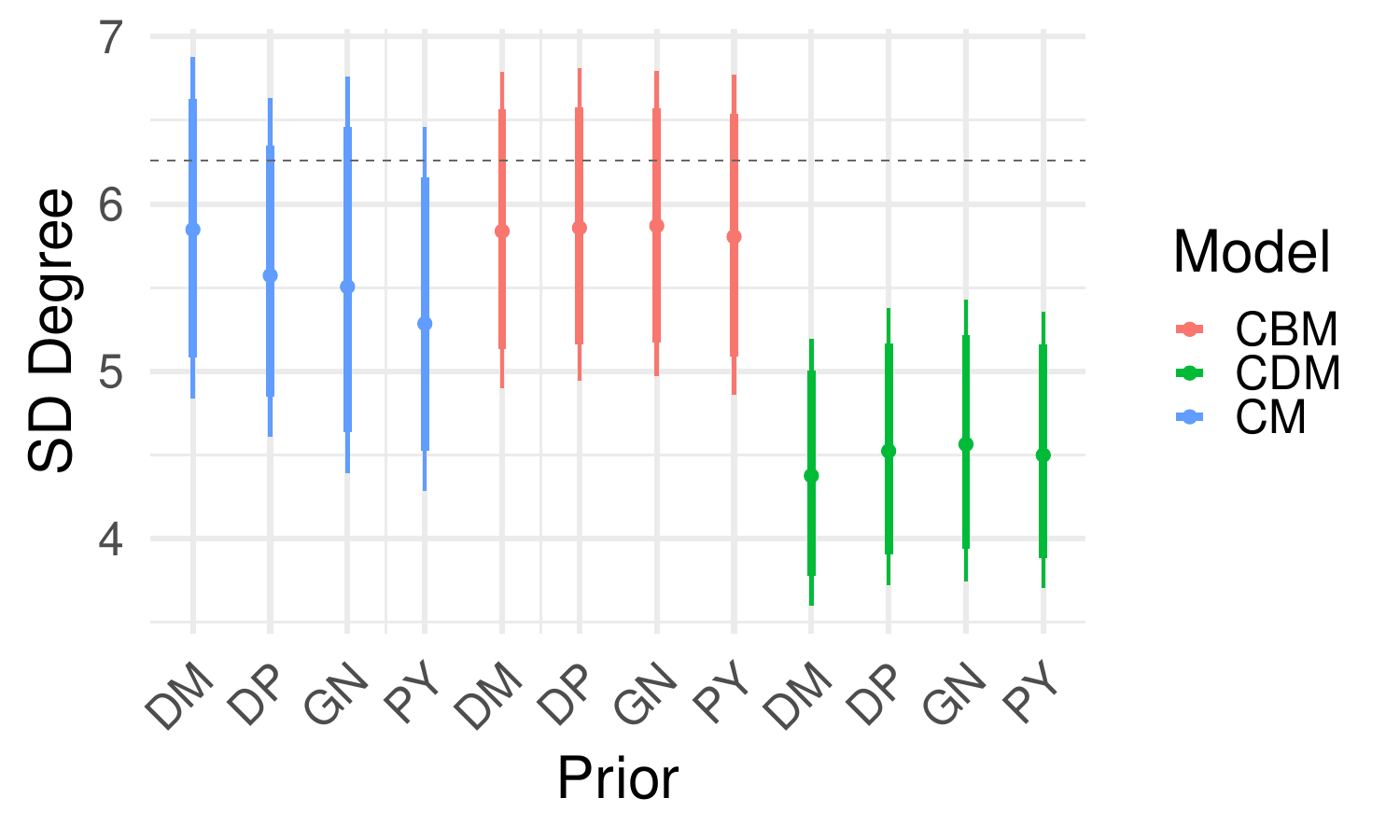}}
    \subfigure[Mean Distance.] {\includegraphics[scale=0.2]{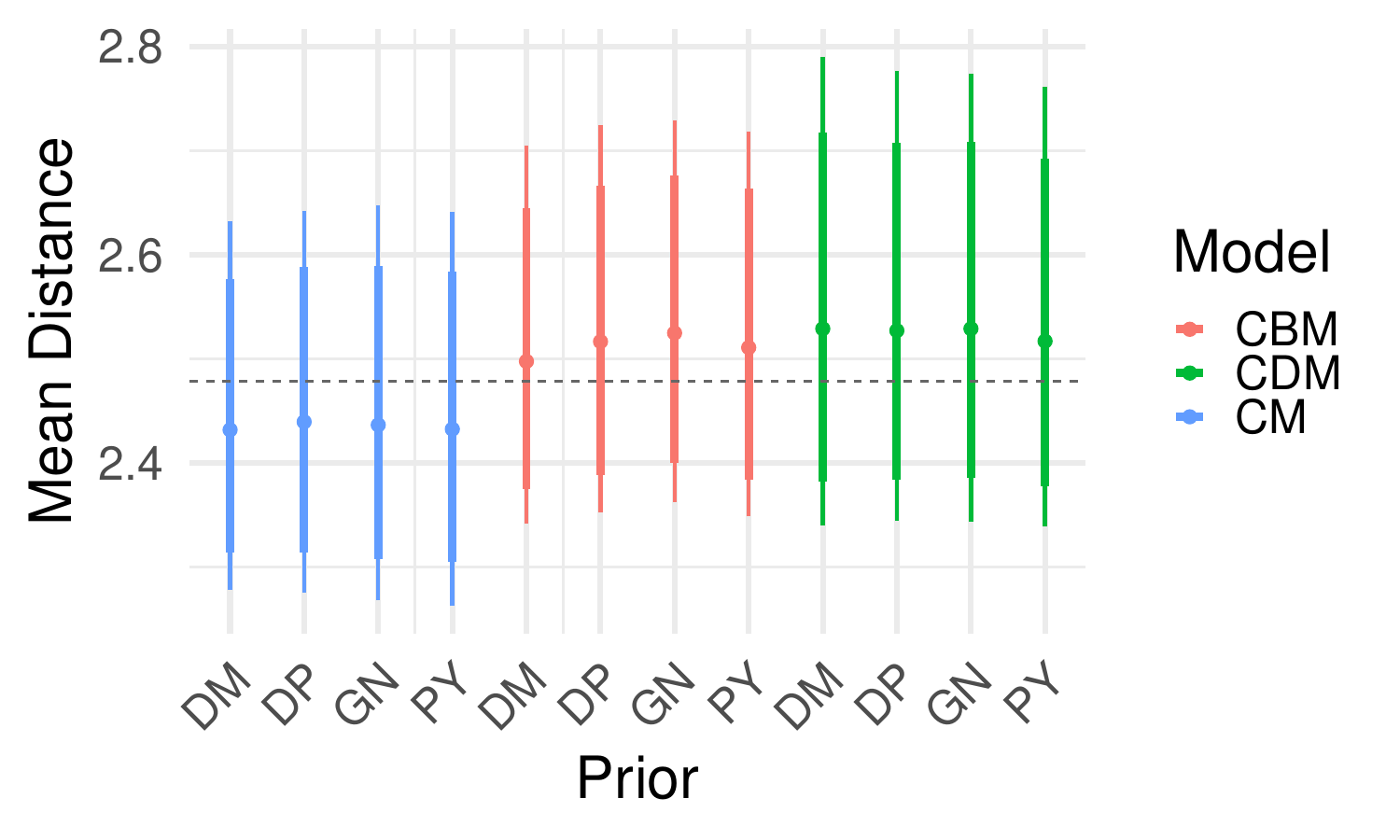}}
    \caption{Posterior predictive checks for key network statistics in the Village dataset.}
    \label{fig:test_statistics_salter}
\end{figure}

\begin{figure}[!htb]
    \centering
    \subfigure[Density.]       {\includegraphics[scale=0.2]{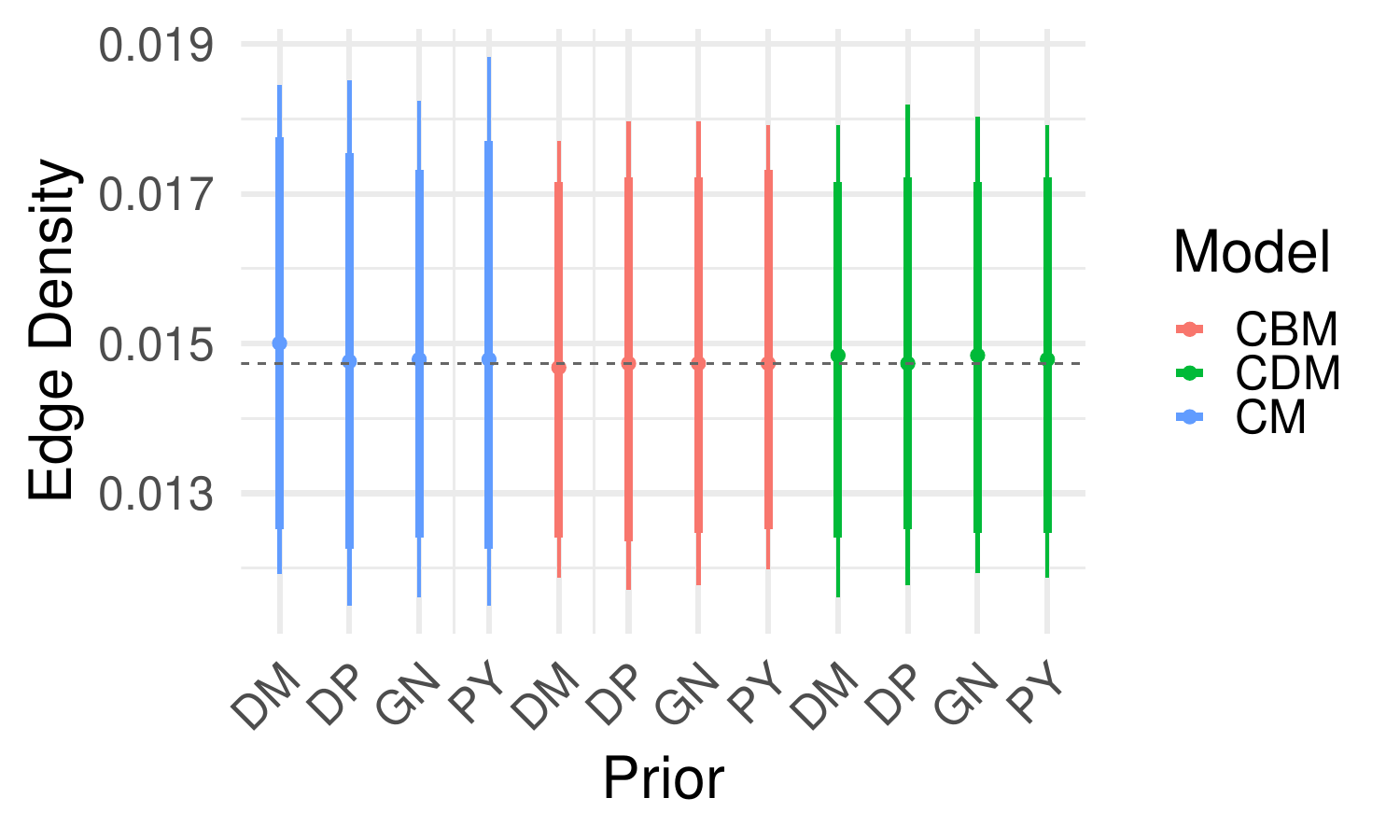}}
    \subfigure[Transitivity.]  {\includegraphics[scale=0.2]{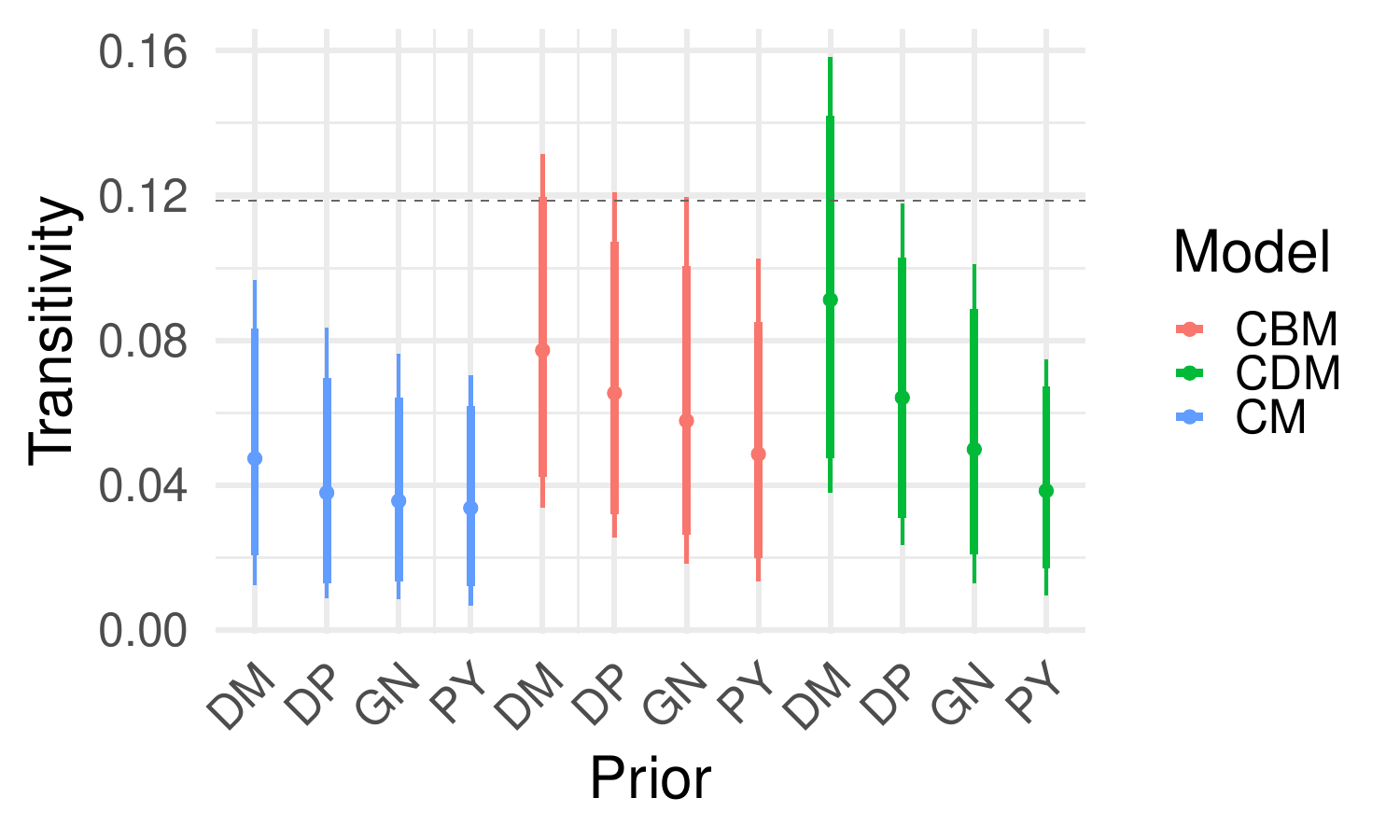}}
    \subfigure[Assortativity.] {\includegraphics[scale=0.2]{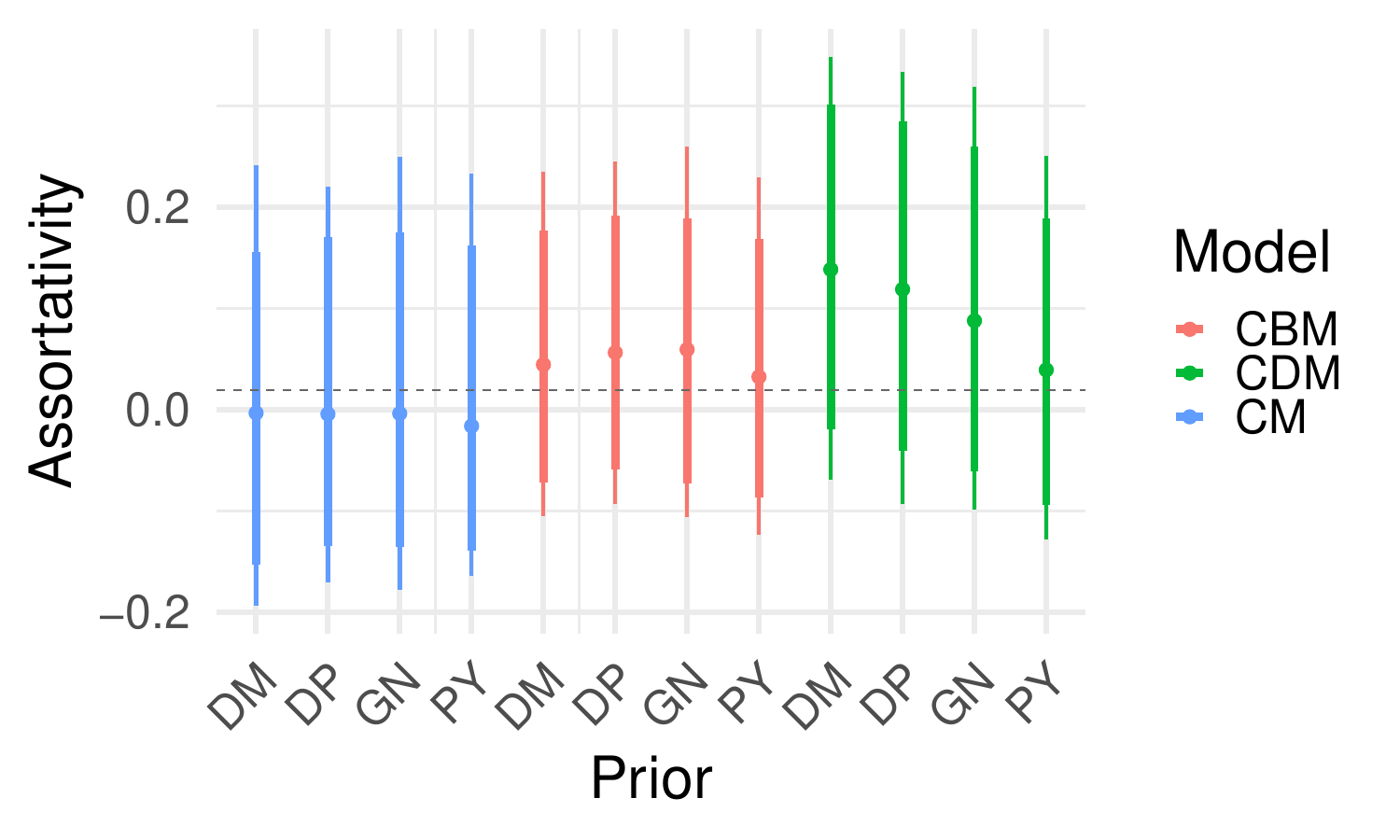}}
    \subfigure[Mean Degree.]   {\includegraphics[scale=0.2]{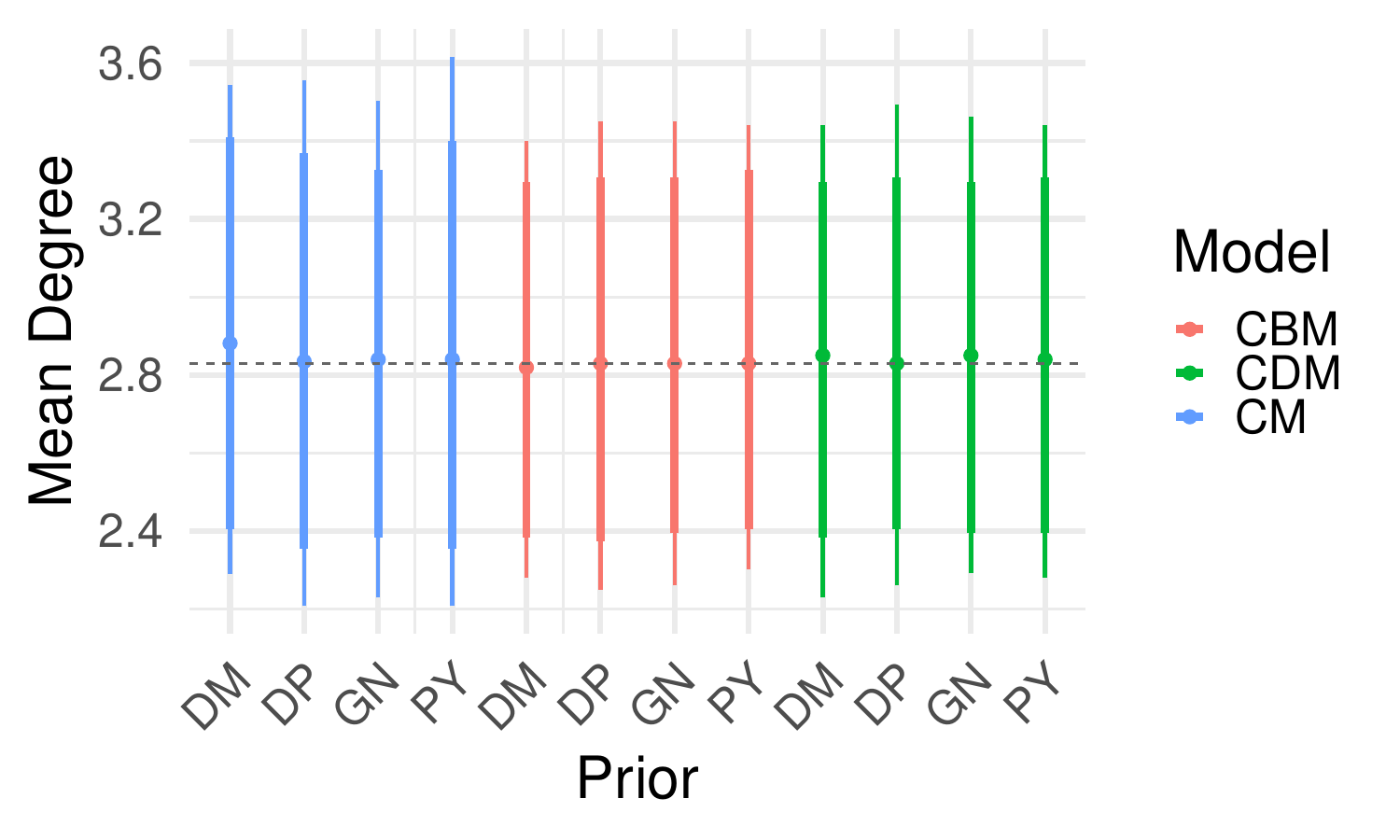}}
    \subfigure[SD Degree.]     {\includegraphics[scale=0.2]{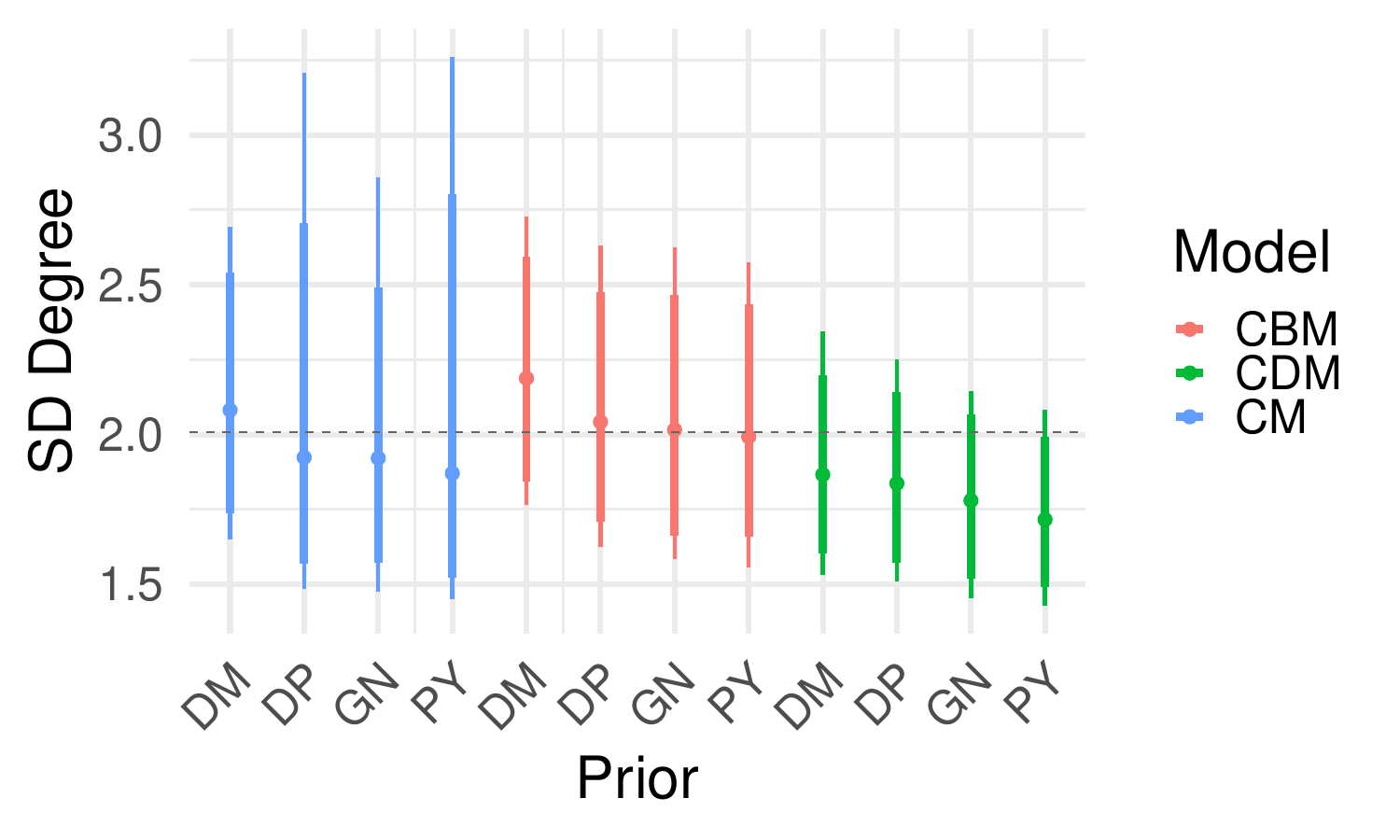}}
    \subfigure[Mean Distance.] {\includegraphics[scale=0.2]{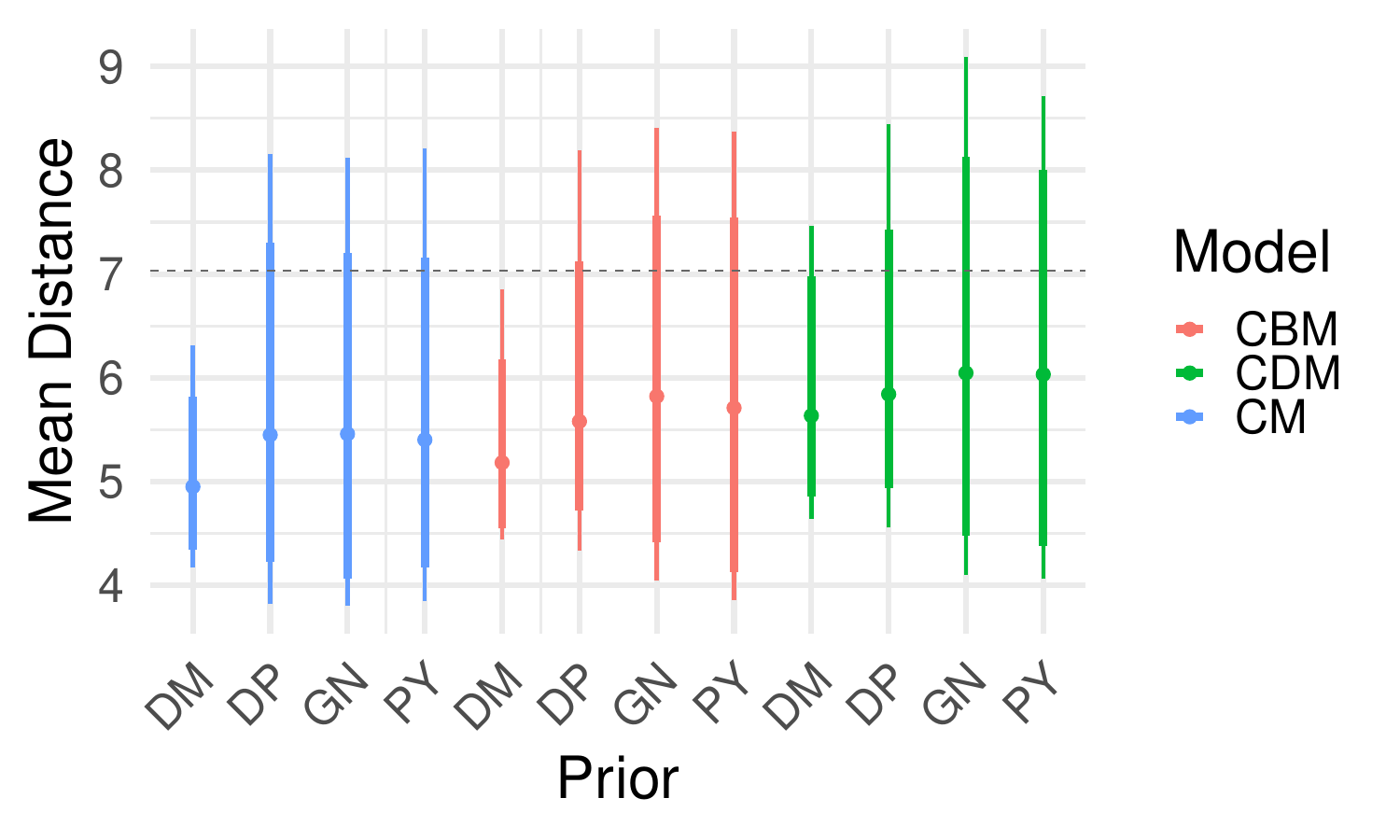}}
    \caption{Posterior predictive checks for key network statistics in the Drug dataset.}
    \label{fig:test_statistics_drug}
\end{figure}

We now evaluate posterior predictive checks for key network statistics in the Village and Drug networks, as shown in Figures~\ref{fig:test_statistics_conflict}, \ref{fig:test_statistics_salter}, and \ref{fig:test_statistics_drug}. Each panel compares the observed statistic (dashed line) against the posterior predictive distributions under different model–prior combinations. Vertical lines represent 95\% and 99\% credible intervals, with point estimates indicating posterior medians. Model classes (CM, CDM, CBM) are distinguished by color, and the x-axis identifies the clustering prior (DM, DP, PY, GN). These diagnostics evaluate how well each model reproduces key structural features of the observed networks.

For the conflict network, posterior predictive analysis confirms that CDM and CBM are very competitive in capturing structural patterns. CBM provides the best overall fit, especially under DP and GN priors, accurately reproducing degree heterogeneity and transitivity, two key features of networks shaped by violence and illicit economies. CDM also improves upon CM but tends to overestimate or underestimate certain statistics depending on the prior. The DM prior performs worst across all models, consistently failing to recover central network characteristics. These findings underscore the advantage of incorporating latent distance or bilinear components at the cluster level, which enhance flexibility and structural realism when modeling complex relational data.

In the Village and Drug networks, CDM and CBM again achieve better recovery of key topological features compared to CM. Both models improve the estimation of transitivity, with CBM showing the closest match in the Village data by better capturing assortativity and degree variability. In the Drug network, the two hybrid models outperform CM in reproducing edge density, mean degree and degree heterogeneity, with CDM being particularly effective in modeling the degree structure. However, CDM tends to overestimate assortativity and both hybrid models slightly overpredict mean geodesic distance, particularly in the Drug network. Overall, the hybrid models demonstrate strong adaptability and robustness across distinct network structures.

\section{Discussion}\label{sec:discussion}

This work introduces a flexible and comprehensive Bayesian framework for network clustering that enhances classical stochastic block models (SBMs) by incorporating low-dimensional latent features at the cluster level. Two new model classes are proposed: the Class-Distance Model (CDM) and the Class-Bilinear Model (CBM). In the CDM, interaction probabilities are governed by the squared Euclidean distances between cluster-specific latent positions, while in the CBM, connections are determined by bilinear interactions among cluster-specific latent vectors. These formulations allow the models to capture both assortative and disassortative patterns, going beyond the constraints of traditional SBMs by encoding latent geometric structures. This work offers a unified and extensible framework for clustering in networks, contributing a full specification of priors, algorithms, and model assessment tools. It serves as both a methodological advancement as well as a practical guide.

All models are fully specified under a Bayesian paradigm. Cluster assignments follow either a finite-dimensional Dirichlet–Multinomial (DM) distribution or nonparametric clustering priors, including the Dirichlet Process (DP), the Pitman–Yor Process (PYP), and the Gnedin–Pitman (GNP) prior. These choices offer flexibility in learning the number of clusters and capturing rich partition structures. Latent variables at the cluster level are given multivariate Gaussian priors, and connectivity effects are modeled through Gaussian distributions with conjugate Inverse-Gamma priors on the variances. This hierarchical specification permits straightforward implementation of posterior inference via Markov Chain Monte Carlo (MCMC), using a combination of collapsed Gibbs samplers and Metropolis steps for non-conjugate parameters. Hyperparameter choices are weakly informative to support data-driven learning while maintaining prior regularization.  Model evaluation includes both synthetic and real-data experiments. The simulation study demonstrates that the proposed latent-cluster models tends to outperform classical SBMs in recovering ground-truth partitions, especially under transitive structures. 

The application to the Colombian conflict network illustrates the value of the proposed framework in capturing complex relational patterns shaped by violence and illicit economies. This network exhibits strong clustering and structural heterogeneity, making it challenging for standard models. The hybrid models CDM and CBM, particularly CBM, outperform the classical SBM in reproducing key features such as transitivity and degree variability. By incorporating latent structure at the cluster level, these models uncover meaningful groupings that reflect territorial dynamics and armed actor influence, offering insights relevant for peacebuilding, drug policy, and targeted intervention strategies.

Several promising directions arise from this work. One natural extension involves adopting alternative latent specifications, such as the eigenmodel, which uses inner products between latent vectors and allows for both positive and negative interaction tendencies. This would enrich the expressiveness of the latent space and potentially capture additional patterns of connectivity. Another avenue is the integration of covariates at the cluster level, either to inform the prior over assignments or to modulate interaction probabilities directly. Such extensions would be especially useful in applications where nodes have rich attributes or are embedded in structured environments. From a computational standpoint, the development of faster inference algorithms is a critical step forward. While the current MCMC samplers offer full posterior inference, scalable approximations such as variational Bayes or stochastic gradient MCMC methods could enable applications to larger networks. Finally, other clustering priors merit exploration. In particular, non-exchangeable or overlapping priors, such as those based on the Normalized Generalized Gamma process or product partition models, could offer new insights into network structures with more complex dependence.

\section*{Statements and declarations}

The authors declare that they have no known competing financial interests or personal relationships that could have appeared to influence the work reported in this article.

During the preparation of this work the authors used ChatGPT-4-turbo in order to improve language and readability. After using this tool, the authors reviewed and edited the content as needed and take full responsibility for the content of the publication.

\bibliography{references.bib}
\bibliographystyle{apalike}




\end{document}